%% file: mechanized.tex
\PassOptionsToPackage{unicode}{hyperref}
\documentclass[sigconf,table]{acmart}%\settopmatter{printfolios=true}
\acmConference[Woodstock '18]{Woodstock '18: ACM Symposium on Neural
  Gaze Detection}{June 03--05, 2018}{Woodstock, NY}
\acmBooktitle{Woodstock '18: ACM Symposium on Neural Gaze Detection,
  June 03--05, 2018, Woodstock, NY}
\acmPrice{15.00}
\acmISBN{978-1-4503-9999-9/18/06}

\setcopyright{none}
\usepackage[T1]{fontenc}
\usepackage{ucs}
\usepackage[utf8x]{inputenc}
\usepackage{autofe}

\usepackage{agda}
\usepackage{catchfilebetweentags}

\DeclareUnicodeCharacter{"2237}{\ensuremath{::}}
\DeclareUnicodeCharacter{"03F1}{\ensuremath{\varrho}}

\usepackage[greek,english]{babel}
\usepackage{cmll}
\usepackage{paralist}
\usepackage{listings}
\usepackage{tikz}
\usetikzlibrary{arrows}

\usepackage{booktabs}   %% For formal tables:
\usepackage{subcaption} %% For complex figures with subfigures/subcaptions
\usepackage{mathpartir}
\input{macros}
\begin{document}

%% Title information
\title[Mechanized Semantics for Session Types]{%
  Intrinsically-Typed Mechanized Semantics for Session Types
}
%\titlenote{with title note}             %% \titlenote is optional;
                                        %% can be repeated if necessary;
                                        %% contents suppressed with 'anonymous'
%\subtitle{Subtitle}                     %% \subtitle is optional
%\subtitlenote{with subtitle note}       %% \subtitlenote is optional;
                                        %% can be repeated if necessary;
                                        %% contents suppressed with 'anonymous'

%% Author information
%% Contents and number of authors suppressed with 'anonymous'.
%% Each author should be introduced by \author, followed by
%% \authornote (optional), \orcid (optional), \affiliation, and
%% \email.
%% An author may have multiple affiliations and/or emails; repeat the
%% appropriate command.
%% Many elements are not rendered, but should be provided for metadata
%% extraction tools.

%% Author with single affiliation.
\author{Peter Thiemann}
% \authornote{with author1 note}          %% \authornote is optional;
                                        %% can be repeated if necessary
\orcid{0000-0002-9000-1239}             %% \orcid is optional
\affiliation{
  % \position{Position1}
  % \department{Informatik}              %% \department is recommended
  \institution{University of Freiburg}            %% \institution is required
  % \streetaddress{Street1 Address1}
  % \city{City1}
  % \state{State1}
  % \postcode{Post-Code1}
  \country{Germany}                    %% \country is recommended
}
\email{thiemann@acm.org}          %% \email is recommended

%% Author with two affiliations and emails.
% \author{First2 Last2}
% \authornote{with author2 note}          %% \authornote is optional;
%                                         %% can be repeated if necessary
% \orcid{nnnn-nnnn-nnnn-nnnn}             %% \orcid is optional
% \affiliation{
%   \position{Position2a}
%   \department{Department2a}             %% \department is recommended
%   \institution{Institution2a}           %% \institution is required
%   \streetaddress{Street2a Address2a}
%   \city{City2a}
%   \state{State2a}
%   \postcode{Post-Code2a}
%   \country{Country2a}                   %% \country is recommended
% }
% \email{first2.last2@inst2a.com}         %% \email is recommended
% \affiliation{
%   \position{Position2b}
%   \department{Department2b}             %% \department is recommended
%   \institution{Institution2b}           %% \institution is required
%   \streetaddress{Street3b Address2b}
%   \city{City2b}
%   \state{State2b}
%   \postcode{Post-Code2b}
%   \country{Country2b}                   %% \country is recommended
% }
% \email{first2.last2@inst2b.org}         %% \email is recommended

%% Abstract
%% Note: \begin{abstract}...\end{abstract} environment must come
%% before \maketitle command
\begin{abstract}
Session types have emerged as a powerful paradigm for structuring communication-based programs. They guarantee type soundness and session fidelity for concurrent programs with sophisticated communication protocols. As type soundness proofs for languages with session types are tedious and technically involved, it is rare to see mechanized soundness proofs for these systems. 

We present an executable intrinsically typed small-step semantics for a realistic functional session type calculus. The calculus includes linearity, recursion, and recursive sessions with subtyping. Asynchronous communication is modeled with an encoding. 

The semantics is implemented in Agda as an intrinsically typed, interruptible CEK machine. This implementation proves type preservation and a particular notion of progress by construction.
\end{abstract}

%% 2012 ACM Computing Classification System (CSS) concepts
%% Generate at 'http://dl.acm.org/ccs/ccs.cfm'.

%% End of generated code

%% Keywords
%% comma separated list
  %% \keywords are mandatory in final camera-ready submission

%% \maketitle
%% Note: \maketitle command must come after title commands, author
%% commands, abstract environment, Computing Classification System
%% environment and commands, and keywords command.
\maketitle

\section{Introduction}
\label{sec:introduction}

Session types are an established paradigm for structured
communication~\cite{Honda1993,HondaVasconcelosKubo1998,DBLP:journals/csur/HuttelLVCCDMPRT16,CastagnaDezaniCiancagliniGiachinoPadovani2009}.
The original form provides a typing discipline for protocols between two parties on  bidirectional, 
heterogeneously typed communication channels. The operations
for session types $S$ are
\begin{itemize}
\item creating a channel;
\item $\syntax{send} : {!T}.S \multimap T \multimap S$, sending a value of type $T$ and
  continuing the protocol according to $S$;
\item  $\syntax{recv} : {?T}.S \multimap T \otimes S$, receiving a value of
  type $T$ and continuing according to $S$;
\item $\syntax{select}_i : (S_1\oplus S_2) \multimap S_i$, internal choice between two continuations;
\item $\syntax{branch} : (S_1 \with S_2) \multimap (S_1 \multimap T) \otimes (S_2 \multimap
  T)$, accepting an external choice between different   continuations;
\item $\syntax{close}: \textbf{end!} \to 1$, closing a channel;
\item $\syntax{wait}: \textbf{end?} \to 1$, waiting for a channel  to be closed.
\end{itemize}
There are synchronous and asynchronous variants
\cite{GayVasconcelos2010-jfp,Wadler2012,DBLP:conf/esop/LindleyM15}, or
variants with additional typing features (e.g., dependent types
\cite{DBLP:conf/fossacs/ToninhoY18}, context-free
\cite{ThiemannVasconcelos2016,DBLP:conf/esop/Padovani17}), as well as
extensions to deal with more than two participants in a protocol
\cite{HondaYoshidaCarbone2008}. They also found their way into
functional and object oriented languages \cite{DBLP:journals/iandc/Dezani-CiancagliniDMY09}.  Generally, session types are inspired
by linear type systems and systems with typestate as channels change
their type at each communication operation and must thus be handled
linearly. Some variants are directly connected to linear logic via the
Curry-Howard correspondence
\cite{CairesPfenning2010,DBLP:conf/esop/ToninhoCP13,Wadler2012}.

What static guarantees do we expect from session types? First, type
preservation --- as for any other type system. Second, session fidelity,
which means that channels are only used according to their stated
protocol/session type. Third, some notion of progress. The strongest
notion of progress is, of course, deadlock freedom, which is
guaranteed by the designs derived from linear logic
\cite{CairesPfenning2010}, but many formulations of session types have
weaker notions of progress.

How does the literature go about proving these properties? Mostly by
careful and tedious manual proofs. To our knowledge, there is only one
session-type related paper that comes with a mechanized proof of type soundness
\cite{DBLP:journals/mscs/GotoJJPR16}, which we discuss in Section~\ref{sec:related}.

\textbf{The main contribution of this paper is the first mechanized proof of the type
  soundness and session fidelity of a realistic, fully-fledged functional session
  type calculus.}
%More precisely:
\begin{itemize}
\item First fully mechanized type preservation proof for a functional
  calculus with session-typed communication.
  
  On top of the functional core (call-by-value PCF with products) our
  formalization covers linear typing, forking processes, creating and 
  closing communication channels, higher-order sending and receiving values,
  selection from and offering of choice, and recursive session types with subtyping.
\item
  First fully mechanized proof of \emph{session fidelity}.

  Session fidelity means that the operations on both ends
  of a channel always agree on the direction of the next transmission on
  the channel and on the type of the transmitted value, if any: an
  operation that sends a value of type $T$ on one end is always
  matched by a receive operation that expects a value of (a supertype
  of) type $T$.

\item The preservation proof on the level of expressions comes
  essentially for free as the calculus is implemented in terms of an
  intrinsically typed, multi-threaded CEK machine.

  For each serious computation step, the CEK machine relinquishes to
  the scheduler, which preserves typing along with some global
  resource invariants.

\item Relative progress is established by construction. The
  scheduler contains a step function that attempts to find a thread in
  which it can perform a computation step. If the step function fails, it has
  detected a deadlock situation (which is to be expected as the
  underlying calculus is not deadlock-free).

\item The focus of the formalization is a sychronous session type
  calculus, but we give a typed encoding of asychronous session types
  in typed ones, which is compatible with our results.
\end{itemize}

% \textbf{Q: Is it entirely straightforward to formalize?}
% \textbf{Q: What will I learn from reading this paper?}

% The main challenges/contributions addressed by our formalization:

% First formalization of a significantly large language with session
% types.

% First one to employ intrinsically typed syntax.

\subsection*{Challenges of the formalization}

Several aspects of the formalization are standard: intrinsically typed
abstract syntax with de Bruijn indexing
\cite{DBLP:conf/sbmf/Wadler18,DBLP:journals/pacmpl/PoulsenRTKV18,DBLP:journals/jar/BentonHKM12}, 
type-indexed values, and gas-driven evaluation \cite{siek13:_type_safet_three_easy_lemmas,DBLP:conf/esop/OwensMKT16, DBLP:conf/popl/AminR17}.
In addition, we address the following key challenges.
\begin{enumerate}
\item Linearity needs to be handled at the syntactic level. The typing of
  expressions must make sure that linearly typed values are neither
  duplicated nor discarded.
\item The multi-threaded CEK machine provides an executable model of
  reductions at the expression level while keeping a clean interface
  to thread-level reduction, which is inspired by free monad
  constructions, and thread scheduling.
  Our model performs only syntactic rearrangements (aka administrative
  reductions) at the expression level and leaves serious execution
  steps to the scheduler at the thread level. This design choice
  pushes the handling of all effects (nontermination, communication,
  nondeterminism) entirely to the thread level, 
  thus simplifying the model for the expression level.
\item Channels need to be managed as a global resource. The problem
  here is that communication over a channel requires a rendezvous
  between two different threads (expressions). These threads must have
  a common means to identify the channel, which cannot be handled
  locally at the level of an expression.
\item Resources need to be handled linearly, in particular when they
  are moved across different semantics entities. We deal with this
  aspect by introducing the concept of a \emph{resource splitting
    tree}, which is a binary tree that maintains linearity.
\item We make a preliminary investigation of adequacy of the CEK
  execution with respect to a small-step semantics. Adequacy does not
  follow from the intrinsic typedness of the interpreter, but requires
  separate proofs (see Section~\ref{sec:adequacy}). 
\end{enumerate}

The framework reported in this paper is implemented in roughly 3000
lines of Agda.
The full code is available in a github repository\footnote{\url{https://github.com/peterthiemann/definitional-session}}.
For pedagogic reasons, 
the code fragments shown in
Sections~\ref{sec:micr-tiny-lang}--\ref{sec:over-struct-interpr} of
this paper are taken from a simplified version with some slight
deviations from the repository version, but nonetheless fully
type checked. The code fragments in Section~\ref{sec:full-picture} are
pretty printed excerpts from the repository code. Throughout, we
assume familiarity with Agda or other similar proof assistants.

\section{A Tiny Language with Session Types}
\label{sec:micr-tiny-lang}

\begin{figure}[tp]
\begin{flushleft}
  \ExecuteMetaData[Typing3.tex]{types}
  \vspace*{-1\baselineskip}
  \ExecuteMetaData[Typing3.tex]{sty-patterns}
  \vspace*{-1\baselineskip}
  \ExecuteMetaData[Typing3.tex]{TCtx}
\end{flushleft}
  \caption{Session types, types, and typing contexts}
  \label{fig:agda-types}
\end{figure}

\begin{figure}[tp]
  \begin{flushleft}
    \ExecuteMetaData[Typing0.tex]{unr}
    \ExecuteMetaData[Typing0.tex]{splitting}
  \end{flushleft}
  \caption{Unrestricted types and context splitting}
  \label{fig:unrestricted-types}
\end{figure}

\begin{figure}[tp]
  \begin{flushleft}
    \ExecuteMetaData[Typing0.tex]{element}
    \ExecuteMetaData[Syntax0.tex]{expr-var}
  \end{flushleft}
  \vspace*{-1\baselineskip}
  \hspace{0.75em}%
  \begin{minipage}{0.975\linewidth}
    \begin{flushleft}
      \ExecuteMetaData[Syntax0.tex]{expr-unit}
      \ExecuteMetaData[Syntax0.tex]{expr-pair}
      \vspace*{-\baselineskip}
      \ExecuteMetaData[Syntax0.tex]{expr-bind}
      \vspace*{-\baselineskip}
      \ExecuteMetaData[Syntax0.tex]{expr-fork}
      \vspace*{-\baselineskip}
      \ExecuteMetaData[Syntax3.tex]{Expr-send}
      \vspace*{-\baselineskip}
      \ExecuteMetaData[Syntax3.tex]{Expr-recv}
    \end{flushleft}
  \end{minipage}
  \caption{Expression typing}
  \label{fig:expression-typing}
\end{figure}

MicroSession is a simplified subset of the session-typed language by
\CITET{Gay and Vasconcelos}{GayVasconcelos2010-jfp}. Nevertheless, it
contains the fundamental
operations of a session-typed language. These operations are channel creation, sending and receiving
values, and closing a channel. To glue these operations together in a
meaningful way, we further require process creation (fork),
introduction and elimination of pairs (the operations to create a
channel and to receive a value return pairs), and a let construct to
sequence operations.  This simplicity enables us to concentrate on
the key ideas, on which the modeling of all further operations on sessions relies.
In Section~\ref{sec:full-picture} we explain what it takes to extend the formalization of
MicroSession to full Synchronous GV.

In this section, we define the type structure and the syntax of
MicroSession. Our syntax definition is intrinsically typed in a way
that enforces linear handling of channels. Variable binding is handled
with de Bruijn indices.

Figure~\ref{fig:agda-types} defines the type structure of MicroSession. Types are unit types, pair
types, and channel types. A channel is described by a session type, which either describes a
transmission (\AgdaInductiveConstructor{STrm}\AgdaSpace{}%
\AgdaBound{d}\AgdaSpace{}%
\AgdaBound{t}\AgdaSpace{}%
\AgdaBound{s}) of a value of type \AgdaBound{t} and continuing at type \AgdaBound{s} or the closing
of a channel (\AgdaInductiveConstructor{SEnd}\AgdaSpace{}%
\AgdaBound{d}). Both types are parameterized by a direction
(\AgdaBound{d} : \AgdaDatatype{Dir}), which
indicates the direction of the data transfer, either sending (\AgdaInductiveConstructor{SND}) or
receiving (\AgdaInductiveConstructor{RCV}). This parameterization
reduces the number of cases to consider when defining duality,
subtyping, and context reduction and proving their properties. 
For convenience, we
introduce the abbreviations 
\AgdaInductiveConstructor{SSend}\AgdaSpace{}%
\AgdaBound{t}\AgdaSpace{}%
\AgdaBound{s},
\AgdaInductiveConstructor{SRecv}\AgdaSpace{}%
\AgdaBound{t}\AgdaSpace{}%
\AgdaBound{s},
\AgdaInductiveConstructor{SEnd!}, and
\AgdaInductiveConstructor{SEnd?}
using Agda's \AgdaKeyword{pattern} feature.

The \AgdaFunction{dual} of a session type describes the protocol on the
other end of the channel (see Fig.~\ref{fig:duality}). It flips the direction of the
communication operations. This operation is involutory as proved in
the lemma \AgdaFunction{dual{-}inv}\AgdaSpace{}%
\AgdaSymbol{:}\AgdaSpace{}%
\AgdaSymbol{∀}\AgdaSpace{}%
\AgdaBound{s}\AgdaSpace{}%
\AgdaSymbol{→}\AgdaSpace{}%
\AgdaFunction{dual}\AgdaSpace{}%
\AgdaSymbol{(}\AgdaFunction{dual}\AgdaSpace{}%
\AgdaBound{s}\AgdaSymbol{)}\AgdaSpace{}%
\AgdaOperator{\AgdaDatatype{≡}}\AgdaSpace{}%
\AgdaBound{s}.\footnote{The type
\AgdaBound{x}\AgdaSpace\AgdaDatatype{≡}\AgdaSpace\AgdaBound{y} denotes 
Martin-L\"of identity of \AgdaBound{x} and \AgdaBound{y} with 
single element \AgdaInductiveConstructor{refl} \cite{MartinLoef1975}.}
Figure~\ref{fig:duality} defines the operation \AgdaFunction{dual} and
proves that it is involutory.
\begin{figure}[tp]
\begin{minipage}[t]{\linewidth}
  \ExecuteMetaData[Typing3.tex]{dual-dir}
  \vspace*{-1\baselineskip}
  \ExecuteMetaData[Typing3.tex]{dual}
  \vspace*{-1\baselineskip}
  \ExecuteMetaData[Typing3.tex]{dual-dir-inv}
  \vspace*{-1\baselineskip}
  \ExecuteMetaData[Typing3.tex]{dual-inv}
\end{minipage}
  \caption{Duality}
  \label{fig:duality}
\end{figure}
We use lists of types as typing contexts (\APhi : \AgdaDatatype{TCtx}).

Figure~\ref{fig:unrestricted-types} contains key notions to formalize linear
handling of resources in the type system
\cite{walker:_subst_type_system,DBLP:conf/lics/CervesatoP96}. 
A ternary context splitting relation\footnote{The definition makes 
  use of Agda's syntactic convention for defining infix operators: the underlines in the
  identifier \AgdaDatatype{\_≜\_∘\_} indicate the positions of the three arguments.  }
\AgdaBound{Φ}\AgdaSpace{}%
\AgdaDatatype{≜}\AgdaSpace{}%
\AgdaBound{Φ₁}\AgdaSpace{}%
\AgdaDatatype{∘}\AgdaSpace{}%
\AgdaBound{Φ₂}
on typing contexts captures contraction. It splits
an incoming context \AgdaBound{Φ} in the conclusion of a rule into two
contexts \AgdaBound{Φ₁} and \AgdaBound{Φ₂} 
for use in the premises. As values of unrestricted type can
be contracted whereas linear ones cannot, there is a predicate
\AgdaDatatype{Unr} that identifies the unrestricted ones: the unit
type is unrestricted and pairs are unrestricted if both components are.
It extends to typing contexts
\APhi{}\AgdaSpace{}%
\AgdaSymbol{:}\AgdaSpace{}%
\AgdaFunction{TCtx} by \AgdaDatatype{All}\AgdaSpace{}%
\AgdaDatatype{Unr}\AgdaSpace{}%
\APhi{}, which means that all entries of
\APhi{} need to satisfy \AgdaDatatype{Unr}.\footnote{Using the library type
  \AgdaDatatype{Data.List.All}.}

Context splitting implements contraction by permitting only bindings
for unrestricted types to end up in both \AgdaBound{Φ₁} and
\AgdaBound{Φ₂} (constructor \AgdaInductiveConstructor{unr}). Linear
types end up either in \AgdaBound{Φ₁} (constructor
\AgdaInductiveConstructor{lft}) or in \AgdaBound{Φ₂} (constructor
\AgdaInductiveConstructor{rgt}).

Figure~\ref{fig:expression-typing} defines the typing of
expressions by means of the type \AgdaDatatype{Expr}\AgdaSpace{}%
\APhi\AgdaSpace{}%
\AgdaBound{t}. It is indexed by the typing context \APhi{} and the
result type \AgdaBound{t} and ensures that its values have type
\AgdaBound{t} in context \APhi. While this setup has become standard \cite{DBLP:journals/pacmpl/PoulsenRTKV18,DBLP:conf/sbmf/Wadler18,DBLP:conf/cpp/Allais0MM17},
there are a few twists in our particular instance.

The syntax is in A-normal form to restrict evaluation contexts to a
sequence of bodies of \texttt{let} expressions. That is, all
constructs except \texttt{let} and \texttt{fork} take only variables
as subterms. In A-normal form, the \texttt{let} expression is the only
place that composes non-trivial computations.  This choice simplifies
the formalization of the semantics as explained in Sec~\ref{sec:commands}.

Variables (\AgdaInductiveConstructor{var}) are represented by de Bruijn
indices into the context. Accessing variables is formalized in the
type \AgdaBound{t}\AgdaSpace{}%
\AgdaDatatype{∈}\AgdaSpace{}%
\AgdaBound{Φ}, which implements weakening only for unrestricted variables. 
If the variable is on top of the context (constructor
\AgdaInductiveConstructor{here}), then all bindings below must be
unrestricted.
If the variable is below the top (constructor
\AgdaInductiveConstructor{there}), then the top binding must be
unrestricted. 

The \AgdaInductiveConstructor{unit} constructor introduces a unit
value if the context is unrestricted.
The \AgdaInductiveConstructor{pair} constructor introduces a pair. It splits the context and
looks up the components. 
The \AgdaInductiveConstructor{letpair} constructor  eliminates a pair and binds the
components in  an expression. It splits the context
between the pair and the expression. 
\AgdaInductiveConstructor{letbind} splits the context between two
expressions and binds the result of the first expression in the
second.
\AgdaInductiveConstructor{fork} takes a unit-typed expression to run in a new thread.
\AgdaInductiveConstructor{new} requires that the context is
unrestricted; it takes a session type and creates a channel of this
type. It returns a pair of dual endpoints for this channel. 
\AgdaInductiveConstructor{close} and \AgdaInductiveConstructor{wait}
close a channel, actively or passively as indicated by the session
types \AgdaInductiveConstructor{SEnd!} and
\AgdaInductiveConstructor{SEnd?}.
\AgdaInductiveConstructor{send} takes two arguments (hence splitting
is needed), a channel of type
\AgdaInductiveConstructor{TChan}\AgdaSpace{}%
\AgdaSymbol{(}\AgdaInductiveConstructor{SSend}\AgdaSpace{}%
\AgdaBound{t}\AgdaSpace{}%
\AgdaBound{s}\AgdaSymbol{)}
which is ready to send a $t$ and a value of type
$t$. It returns the continuation channel of type $s$.
\AgdaInductiveConstructor{recv} takes a channel which is ready to
receive a $t$ and returns a pair consisting of the continuation
channel and the received value. This pair type is linear because it
contains a channel type.

\input{fig-microsession-example}
% \begin{figure}[tp]
%     \ExecuteMetaData[Examples0.tex]{ex1}
%     \caption{Example term}
%     \label{fig:micro-session-example}
% \end{figure}

%%% TBC

Figure~\ref{fig:micro-session-example} contains a well-formed example
term. Fig.~\ref{fig:example-in-math} shows it in some 
concrete syntax whereas Fig.~\ref{fig:example-in-agda} contains its
Agda rendition. The code creates a session of type
\AgdaInductiveConstructor{SEnd!}  which yields a pair of channel
ends. Next, it decomposes the pair into the channel ends of type
\AgdaInductiveConstructor{SEnd!} and its dual
\AgdaInductiveConstructor{SEnd?}, respectively, forks a thread which
closes one channel end, and keeps the other end in the main thread to
wait on it.

\section{Operational Semantics}
\label{sec:oper-semant}

The semantics has two layers as common in
functional calculi with session types. The top layer 
is a process calculus which operates on a thread pool. Each single
thread corresponds to an expression. A scheduler selects the next
thread to execute.  The bottom layer deals with
the reduction of single expressions.

The expression semantics is modeled by an extended CEK machine
\cite{FelleisenFriedman1986}. The state of a CEK machine
comprises an expression $e$ (aka control), an environment $\varrho$, and a
continuation $\kappa$. This machine is typed so that the environment
contains values according to the type environment for $e$ and the
input type of the continuation matches the type of $e$.
The machine is extended to implement
``stuttering execution''. For any serious activity, it suspends itself to a continuation and renders control
to the scheduler that administers computation on the thread level. In
particular, execution is suspended for every operation that may not
terminate (i.e., function application) or that requires
interaction with the scheduler (i.e., thread and channel creation) or
with other processes (e.g., sending and receiving values).

Expression level execution is driven by a function \AgdaFunction{decompose} that takes a
CEK machine state and returns a continuation along with the operation
that needs to be performed by the scheduler as well as arguments for
this operation. 
In a first approximation the type of the
\AgdaFunction{decompose} function is as
follows.\footnote{Section~\ref{sec:real-type-decompose} discusses the
  full type for \AgdaFunction{decompose}.}
\begin{code}%
\>[0]\AgdaFunction{decompose}\AgdaSpace{}%
\AgdaSymbol{:}\AgdaSpace{}%
\AgdaSymbol{(}\AgdaBound{split}\AgdaSpace{}%
\AgdaSymbol{:}\AgdaSpace{}%
\AgdaBound{Φ}\AgdaSpace{}%
\AgdaDatatype{≜}\AgdaSpace{}%
\AgdaBound{Φ₁}\AgdaSpace{}%
\AgdaDatatype{∘}\AgdaSpace{}%
\AgdaBound{Φ₂}\AgdaSymbol{)}\AgdaSpace{}%
\\
\>[2][@{}l@{\AgdaIndent{0}}]%
\>[6]
\AgdaSymbol{→}\AgdaSpace{}%
\AgdaSymbol{(}\AgdaBound{e}\AgdaSpace{}%
\AgdaSymbol{:}\AgdaSpace{}%
\AgdaDatatype{Expr}\AgdaSpace{}%
\APhi\AgdaBound{₁}\AgdaSpace{}%
\AgdaBound{t}\AgdaSymbol{)}\AgdaSpace%
\AgdaSymbol{→}\AgdaSpace{}%
\AgdaSymbol{(}\AgdaBound{ϱ}\AgdaSpace{}%
\AgdaSymbol{:}\AgdaSpace{}%
\AgdaDatatype{VEnv}\AgdaSpace{}%
%\AgdaBound{G₁}\AgdaSpace{}%
\APhi\AgdaBound{₁}\AgdaSymbol{)}\AgdaSpace%
\AgdaSymbol{→}\AgdaSpace{}%
\AgdaSymbol{(}\AgdaBound{κ}\AgdaSpace{}%
\AgdaSymbol{:}\AgdaSpace{}%
\AgdaDatatype{Cont}\AgdaSpace{}%
%\AgdaBound{G₂}\AgdaSpace{}%
\APhi\AgdaBound{₂}\AgdaSpace{}%
\AgdaBound{t}\AgdaSymbol{)}\AgdaSpace%
\\
\>[2][@{}l@{\AgdaIndent{0}}]%
\>[6]
\AgdaSymbol{→}\AgdaSpace{}%
\AgdaDatatype{Command}\AgdaSpace{}%
%\AgdaBound{G}\<%
\end{code}
The types of the expression \AgdaBound{e} and the value environment \Arho{} are linked via the
type environment \APhi\AgdaBound{₁}, where \AgdaDatatype{VEnv}\AgdaSpace{}%
\APhi\AgdaBound{₁} contains values as specified by the type
environment. The type \AgdaBound{t}  serves as the argument type of the
continuation \Akappa{} as well as the result type of the
expression. The continuation is represented by a closure consisting of
an evaluation context and a value environment. It 
is therefore indexed by type environment \APhi\AgdaBound{₂} which provides the typing of
the variables captured in the closures. The two type environments are split off a single
parent environment \APhi{} to correctly model linear types as
explained in Section~\ref{sec:micr-tiny-lang}. 

% The \AgdaFunction{decompose} function is a decomposition function in the
% sense of Danvy and others \cite{DBLP:conf/pepm/DanvyJZ11}. It splits
% an expression into a potential redex and an evaluation context,
% reified in the type \AgdaDatatype{Command}.
The intention is that the CEK execution ``gives up'' as soon as some
side effect is to be executed. It does so by issuing a
\AgdaDatatype{Command}, which instructs the top-level scheduler to
perform the effect on behalf of the thread. Each
\AgdaDatatype{Command} contains a continuation which accepts the value
resulting from executing the effect as input.

The function \AgdaFunction{decompose} connects the static semantics
with the dynamic semantics. While expressions are indexed by objects
of the \emph{static semantics} (i.e., type environments and types),
(value) environments and continuations are objects of the
\emph{dynamic semantics} and will be indexed with run-time
\emph{resources}. In our case, resources are the endpoints of
communication channels as explained in
Section~\ref{sec:glob-sess-cont}. An endpoint is a special case of a
value, values are contained in environments, and every continuation
contains an environment. Hence, resources are expressed as additional
indices of the types of these objects as shown in
Section~\ref{sec:real-type-decompose}.

The process level consists of expression
processes, parallel execution of processes, and restriction to
introduce a new session typed channel. We leave omit its description
%in the Appendix~\ref{sec:processes}
because we keep processes in a normal form where all restrictions are pushed to the
top level and the expression processes are collected in a thread pool.
The scheduler discards unit processes and enforces commutativity and
associativity.

\section{Resources}
\label{sec:over-struct-interpr}

Working towards resources, we start with defining the values stored in
the environments of the machine. The
main complication arises in devising the representation of channel endpoints.
\begin{itemize}
\item A channel endpoint is a global resource, so it must be \textbf{defined outside} the
  currently executing thread / expression.
\item As two threads communicate through the endpoints of the same channel, there must be a
  \textbf{global naming scheme} for channels across all currently executing threads. 
\item It must be possible to detect a rendezvous of two threads. For example, if
  one thread wants to close a particular channel, we need to
  \textbf{match} it with another thread that
  waits on \textbf{the other end of the same channel}.
\end{itemize}
To this end, we define two entities. A \textbf{global session context} $G$
that corresponds to the collection of all restrictions and a
\textbf{thread pool} that contains representations of all expression processes.
% \begin{align*}
%   P &::= \PExp e \mid \PPar P P \mid \PRes c s P
% \end{align*}

\subsection{Global Session Context}
\label{sec:glob-sess-cont}

The global session context represents the system-wide view of all
channels and their current state. The crucial step is called
\emph{resource splitting} which break this
view down to individual threads and all the way to individual values
while guaranteeing uniqueness of reference and consistency with the
system view. Uniqueness means that each channel end is only referenced
in one value in the current state. As values are related to
environments, continuations, threads, and the threadpool, splittings
naturally form a tree structure, the \emph{splitting tree}. 
Moreover, these conditions (and thus the splitting tree) need to be
maintained throughout the execution.  

\begin{figure}[tp]
  \begin{subfigure}[b]{\linewidth}
    %\hspace*{-2em}
    \begin{minipage}[t]{1.0\linewidth}
    %   \begin{minipage}[t]{0.38\linewidth}
    \ExecuteMetaData[Global0.tex]{SCtx}
    % \end{minipage}
    % \begin{minipage}[t]{0.58\linewidth}
    \vspace*{-1.5\baselineskip} \ExecuteMetaData[Global0.tex]{Inactive}
    % \end{minipage}

  \end{minipage}
    \caption{Global session context and the \AgdaDatatype{Inactive} relation}
    \label{fig:global-session-context}
  \end{subfigure}
  \\ % \quad
  \begin{subfigure}[b]{\linewidth}
    \begin{center}
      \begin{tabular}{lll}
        \AgdaInductiveConstructor{nothing}
        & channel is not available
        & \ChanNothing
        \\
        \AgdaInductiveConstructor{just} (\AgdaFunction{s}, \AgdaInductiveConstructor{POS})
        & positive end is available
        & \ChanPos
        \\
        \AgdaInductiveConstructor{just} (\AgdaFunction{s}, \AgdaInductiveConstructor{NEG})
        &negative end is available
        & \ChanNeg \\
        \AgdaInductiveConstructor{just} (\AgdaFunction{s}, \AgdaInductiveConstructor{POSNEG})
        &both ends are available
        & \ChanPosNeg
      \end{tabular}
    \end{center}
    \caption{Meaning of an entry of type \AgdaDatatype{Session} with graphical encoding. The value
      \AgdaFunction{s} specifies the session type of the positive
      channel end.}
    \label{fig:meaning-of-an-entry}
  \end{subfigure}
  \caption{Global session context}
  \label{fig:super-sctx}
\end{figure}
Figure~\ref{fig:global-session-context} contains the definition of the global session
context by the type \AgdaDatatype{SCtx}, ranged over by $G$. Working backwards,
a global session context is a list. Each entry of the list describes the state of one channel (type
\AgdaDatatype{Session}) in terms of a session type and a polarity of type
\AgdaDatatype{PosNeg}. It contains an entry for every channel that has been
created at some earlier point during execution. Figure~\ref{fig:meaning-of-an-entry} explains the 
meaning of an entry and defines a graphical representation for it. If a channel
is not available at the top level, then it is closed. In general, unavailability
of a channel may also mean that the channel is allocated to a different semantic entity.
A session context is \AgdaDatatype{Inactive} if all its entries 
are unavailable, i.e., equal to 
\AgdaInductiveConstructor{nothing}.\footnote{\AgdaFunction{Is-nothing} is a
  library predicate for the \AgdaDatatype{Maybe} type.}

% A channel is inactive
%  polarities. As already
% mentioned, the model employs a scheduler at the level of processes. This scheduler
% maintains a session context \AgdaBound{G} as a list of resources of type
% \AgdaDatatype{Session}. Each list 
% entry describes a channel / session that has been created at some earlier point during execution;
% such a channel may have been closed already. An entry declares ownership for each end of
% the channel separately. 

To manage resources, each semantic entity that may depend on
the availability of a resource is indexed by a session context
\AgdaBound{G} which makes available \textbf{exactly} the resources used by this entity.
These semantic entities are (in order of containment)
\begin{inparaenum}
\item the thread pool (i.e., a process in normal form),
\item expression threads, 
\item value environments, 
\item values.
\end{inparaenum}

It is an invariant of our encoding that
all these entities refer to a session context of the same length as the global context, so that a channel can
be globally identified by its de Bruijn index into the context.
The thread pool is indexed by the root context from which all other contexts are split
off. Conceptually, the allocation of resources to semantic entities is described
by a single binary tree rooted in the global session context. Each inner node of
the tree corresponds to a  split operation and the leaves correspond to
(primitive) values. Each semantic action maintains this tree.
The upcoming definition of resource splitting guarantees linear handling of all
resources. 

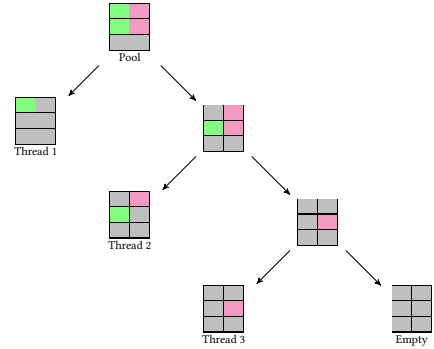
\begin{figure}[tp]
  \centering \scalebox{0.5}{\input{pic-resource-splitting-threads}}
  \caption{Resource splitting for threads}
  \label{fig:resource-splitting-threads}
\end{figure}
For example, Figure~\ref{fig:resource-splitting-threads} shows the splitting
tree for a thread
pool\footnote{Section~\ref{sec:scheduler} contains the definition of the thread pool
  type.} which has two channels available and a third channel that has been closed already. The
channel ends are split among three threads where Thread1 and Thread3 each obtain one end
of a channel, and Thread2 obtains two ends of different channels. In the end, at the Empty
thread pool, all resources must have been distributed to the threads, and we say the
remaining resource context is \textbf{inactive}.

Exactly the same kind of resource splitting occurs in value environments and also in
values themselves.
%(see appendix Section~\ref{sec:resource-splitting}).

\begin{figure}[tp]
  \begin{flushleft}
    \ExecuteMetaData[Global0.tex]{RSplit} 
    \ExecuteMetaData[Global0.tex]{SSplit2}
  \end{flushleft}
  \caption{Resource splitting}
  \label{fig:splitting-session-context}
\end{figure}
Given this intuition, we define resource splitting in
Figure~\ref{fig:splitting-session-context}. Relation \AgdaDatatype{RSplit} governs
splitting of a single entry. Unavailable resources cannot be split
(\AgdaInductiveConstructor{rs-both}); all resources can go exclusively to the left or to
the right child context (\AgdaInductiveConstructor{rs-left},
\AgdaInductiveConstructor{rs-right}); a \AgdaInductiveConstructor{POSNEG} resource stands
for both channel ends and can be split in two ways: either the
\AgdaInductiveConstructor{POS} end goes left and the \AgdaInductiveConstructor{NEG} end
goes right (\AgdaInductiveConstructor{rs-posneg}) or vice versa
(\AgdaInductiveConstructor{rs-negpos}). Relation \AgdaDatatype{SSplit} lifts
\AgdaDatatype{RSplit} to lists of entries in the obvious way.

A splitting
\AgdaBound{ssp-$G=G_1\circ G_2$}\AgdaSpace:\AgdaSpace\AgdaDatatype{SSplit}\AgdaSpace{$G$}\AgdaSpace{$G_1$}\AgdaSpace{$G_2$} 
may be viewed as a separating conjunction $G_1 \ast G_2$~\cite{Reynolds2002}.

\begin{figure}[tp]
  \begin{subfigure}[t]{0.48\linewidth}
    \begin{center}
      \input{pic-ssplit-compose-left0}
    \end{center}
    \caption{Before}
    \label{fig:ssplit-before0}
  \end{subfigure}
  \begin{subfigure}[t]{0.48\linewidth}
    \begin{center}
      \input{pic-ssplit-compose-right0}
    \end{center}
    \caption{After}
    \label{fig:ssplit-after0}
  \end{subfigure}
  \caption{Action of \AgdaFunction{ssplit-compose}}
  \label{fig:ssplit-compose}
\end{figure}
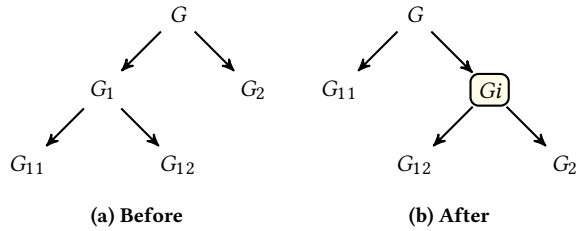
Semantic actions require restructurings of the
splitting tree, which are reminiscent of rotations on balanced trees.
Here is an example for such a restructuring lemma. 
\ExecuteMetaData[Global0.tex]{ssplit-compose}
It transforms the splitting tree in
Figure~\ref{fig:ssplit-before0} to the one in
Figure~\ref{fig:ssplit-after0}, highlighting the new intermediate session context \AgdaBound{Gi}. The proof is straightforward, but tedious.

\subsection{Values and Value Environments}
\label{sec:values}

\begin{figure}[tp]
  \begin{flushleft}
    \ExecuteMetaData[Typing0.tex]{Polarity}
    \ExecuteMetaData[Session0.tex]{ChannelRef}  
  \end{flushleft}
  \caption{Polarities and channel references}
  \label{fig:polarities-and-channel-references}
\end{figure}
\begin{figure}[tp]
  \hspace*{-2em}\begin{minipage}[t]{0.48\linewidth}
    \ExecuteMetaData[Session0.tex]{Val}
  \end{minipage}
  \begin{minipage}[t]{0.38\linewidth}
    \ExecuteMetaData[Session0.tex]{VEnv}
  \end{minipage}
  \caption{Values and value environments}
  \label{fig:values-environments}
\end{figure}
We continue bottom up by defining polarities and channel references in
Fig.~\ref{fig:polarities-and-channel-references}.   The latter are
augmented de Bruijn indices that describe channel values.
There are two ``zeroes'' in type \AgdaDatatype{ChannelRef}, \AgdaInductiveConstructor{here-pos} and
\AgdaInductiveConstructor{here-neg}, depending on whether we found the positive or
negative end of a channel indicated by its polarity argument. For the
positive end, the session type \AgdaBound{s} is copied from the
context. For the negative end, the index is the dual of the session
type in the context. In both cases, the remaining 
context must be inactive. The ``successor'' \AgdaInductiveConstructor{there} skips over
unavailable channels. 

Values and value environments (Figure~\ref{fig:values-environments})
are rather unsurprising at this point.  
 A unit value (like any other primitive value) requires an inactive context.
A pair value splits its resources among the components.
A channel value consists of a \AgdaBound{po} component, which indicates whether this
value is the positive or negative end of the channel, and a channel reference \AgdaBound{cr}
that makes the actual connection to the session context \AgdaBound{G}.

The empty environment \AgdaInductiveConstructor{vnil} requires an inactive session
context. A non-empty environment \AgdaInductiveConstructor{vcons} splits the session
context between the value and the rest of the environment.

\subsection{Decomposing Expressions}
\label{sec:real-type-decompose}
\begin{figure}[tp]
  \ExecuteMetaData[Session0.tex]{Cont}
  \caption{Continuations / Evaluation contexts}
  \label{fig:agda-continuations}
\end{figure}

This subsection explains the types involved in the evaluation of
expressions and processes and how expression
evaluation and the scheduler work together.  The next subsection
discusses the treatment of several illustrative cases across the levels.
We start with the full type of \AgdaFunction{decompose}.
\ExecuteMetaData[Session0.tex]{decompose}
The \AgdaFunction{decompose} function is indexed by a type
environment \APhi{} and by a global session context \AgdaBound{G}, which are both split
using their respective splitting relation. The
splitting of the session context refines the splitting of the typing
environment:  A value bound to a variable may be a (nested) pair that contains several
resources (channel ends). The refinement arises because
the typing environment manages variable bindings, whereas 
the session context keeps track of single resources.

One part of the resources (indexed by \AgdaBound{G₁}) is referenced in the environment
\AgdaBound{ϱ} and manipulated by the expression \AgdaBound{e}. The other part (indexed by
\AgdaBound{G₂}) is referenced and manipulated by the continuation \Akappa{}. A
continuation (see Figure~\ref{fig:agda-continuations}) is a syntactic
object of type
\AgdaDatatype{Cont}\AgdaSpace{}%
\AgdaBound{G}\AgdaSpace{}%
\AgdaBound{Φ}\AgdaSpace{}%
\AgdaBound{t}
where context \AgdaBound{G} describes the resources held,
\AgdaBound{Φ} is the typing for its variables, and
\AgdaBound{t} is its argument type.
It is
either \AgdaInductiveConstructor{halt}, in which case an inactive session context
indicates that all resources have been used up, or a \AgdaInductiveConstructor{bind}
continuation, which is a frozen call to \AgdaFunction{decompose}
consisting of an expression, an environment, and another continuation
along with splittings for variables and resources.

The \AgdaFunction{decompose} function returns a value of type
\AgdaDatatype{Command}\AgdaSpace\AgdaBound{G}, where the session context \AgdaBound{G}
summarizes all resources contained in the command, i.e., in the values
and continuations contained in it. A command is a request to the scheduler to perform
an operation on behalf of the thread. It contains all parameters of the operation and a
continuation to resume the thread. The next subsection
discusses illustrative cases for commands.

An important point of this setup is that \textbf{all values}
manipulated during execution \textbf{are exposed in top-level data structures}. All closures (in
continuations and otherwise) are syntactic and can be inspected and modified at run time
by traversing the respective data structure. This property is preserved up to the
top-level where the scheduler processes the thread pool. We need this
property in Section~\ref{sec:channel-creation} when dealing with
channel creation.

\subsection{Issuing Commands}
\label{sec:commands}

This subsection discusses selected cases of the \AgdaFunction{decompose} function and
the corresponding cases of the \AgdaDatatype{Command} type side by
side. So there is a mix of code from the function definition of  \AgdaFunction{decompose} and
constructor definitions for \AgdaDatatype{Command}. 

The \AgdaFunction{decompose} function follows the evaluation context.
To process an expression
\AgdaInductiveConstructor{letbind}\AgdaSpace\AgdaBound{sp}\AgdaSpace\AgdaBound{e₁}\AgdaSpace\AgdaBound{e₂},
the interpreter splits the value environment according to the split \AgdaBound{sp} of the
typing environment, and constructs a new continuation for \AgdaBound{e₁} by using
\AgdaInductiveConstructor{bind} to compose the frame for \AgdaBound{e₂} with the
continuation \AgdaBound{κ}.

At this point our choice of A-normal form pays off. Without
A-normal form, we would have to perform the decomposition of
\AgdaInductiveConstructor{letbind} that leads to the construction of
the \AgdaInductiveConstructor{bind} continuation 
for each expression with more than one subexpression.
\ExecuteMetaData[Session0.tex]{decompose-bind}
The function \AgdaFunction{split-env} splits the incoming environment
\AgdaBound{ϱ} according to the splitting \AgdaBound{sp},
\AgdaFunction{split-rotate} performs a splitting rotation for typing
environments, and the action of \AgdaFunction{ssplit-compose} is explained in
Figure~\ref{fig:ssplit-compose}. We omit their definitions from the paper.

If \AgdaFunction{decompose} finds a value expression (e.g., a variable, a unit,
or a pair), it constructs the value and arranges to invoke the continuation.
For illustration, we consider the case for unit. Execution of such a value
expression creates a \AgdaInductiveConstructor{Ready} command. It
represents a pending computation by a value and a
continuation. This command instructs the scheduler to apply the
continuation to the value. 
\ExecuteMetaData[Session0.tex]{Command-Ready}
\vspace*{-\baselineskip}
\ExecuteMetaData[Session0.tex]{decompose-unit}
The argument to \AgdaInductiveConstructor{VUnit} is the proof that the
value environment \AgdaBound{ϱ} contains no resources, which follows from the fact that the
typing environment is unrestricted as evidenced by
\AgdaBound{unr{-}Φ}. Lemma \AgdaFunction{inactive-venv}
provides this proof.

The other value expressions follow the same pattern. For example, the
implementation of variable access fetches 
the value from the environment using \AgdaFunction{access} and then requires some
type manipulation using Agda's \AgdaKeyword{rewrite} facility, which
we leave uncommented. As this pattern occurs for all one argument
operations, we abstract it in a function \AgdaFunction{decompose-one}
(type elided).
\ExecuteMetaData[Session0.tex]{decompose-var}
\ExecuteMetaData[Session0.tex]{decompose-one}

The remaining operations (\AgdaInductiveConstructor{fork},
\AgdaInductiveConstructor{new}, \AgdaInductiveConstructor{close},
\AgdaInductiveConstructor{wait}, \AgdaInductiveConstructor{recv},
\AgdaInductiveConstructor{send}) just yield an appropriate
\AgdaDatatype{Command}. 

% \ExecuteMetaData[Session0.tex]{Command}
% For each value of type \AgdaDatatype{Command}\AgdaSpace\AgdaBound{G}, the session context
% \AgdaBound{G} indicates which resources are held by the command.
% A command represents an executing thread in the thread pool.
% It interfaces the thread to the scheduler.

% \ExecuteMetaData[Session0.tex]{Command-Halt}
% The \AgdaInductiveConstructor{Halt} command arises from applying the
% \AgdaInductiveConstructor{halt} continuation to a value. It can be proven that this value
% is unrestricted. Hence, the scheduler will drop a halted thread.

\ExecuteMetaData[Session0.tex]{Command-Fork}  
The \AgdaInductiveConstructor{Fork} command contains two continuations
(thunks, really), one for the 
newly created thread and the other for continuing the existing thread. The scheduler
applies each thunk to the unit value. Its construction in
\AgdaFunction{decompose} is a tedious exercise in resource shuffling
and hence omitted.

\ExecuteMetaData[Session0.tex]{Command-New}
\vspace*{-1\baselineskip}
\ExecuteMetaData[Session0.tex]{decompose-new}
The \AgdaInductiveConstructor{New} command asks the scheduler to create a new channel. The
\AgdaBound{s} component specifies the session type of the channel and the continuation
component \Akappa{} expects a pair of channel endpoints with types \AgdaBound{s} and
\AgdaFunction{dual}\AgdaSpace\AgdaBound{s}.

\ExecuteMetaData[Session0.tex]{Command-Close}
\vspace*{-1\baselineskip}
\ExecuteMetaData[Session0.tex]{decompose-close}
The \AgdaInductiveConstructor{Close} command is just another one
argument expression and can be treated like the variable access. The
definition and processing of the 
\AgdaInductiveConstructor{Wait} and \AgdaInductiveConstructor{Recv}
commands are analogous. Decomposition of the
\AgdaInductiveConstructor{Send} command is slightly more involved
because the \AgdaInductiveConstructor{send} expression takes a
splitting, a channel value, and a value to send. Hence, before
accessing the environment $\varrho$ as with
\AgdaInductiveConstructor{close}, we first have to split the incoming
environment using \AgdaFunction{split-env}. Each part of the
environment is treated as shown in \AgdaFunction{decompose-one}.

\subsection{Executing Commands in the Thread Pool}
\label{sec:scheduler}
\begin{figure}[tp]
  \begin{flushleft}
  \ExecuteMetaData[Schedule3.tex]{scheduler-event}
  \ExecuteMetaData[Schedule3.tex]{scheduler-types}
  \end{flushleft}
  \caption{Auxiliary types for the scheduler}
  \label{fig:types-for-the-scheduler}
\end{figure}
\begin{figure}[tp]
  \begin{flushleft}
    \ExecuteMetaData[TypeReduction3.tex]{type-reduction-internal}
    \vspace*{-\baselineskip}
    \ExecuteMetaData[TypeReduction3.tex]{type-reduction}
    \vspace*{-\baselineskip}
    \ExecuteMetaData[Schedule3.tex]{scheduler-nextpool}
  \end{flushleft}
  \caption{Context reduction}
  \label{fig:context-reduction}
\end{figure}

The scheduler actually drives the execution. It
operates on the  thread pool, which is represented by a list of commands.

\ExecuteMetaData[Session0.tex]{ThreadPool}
The type of
the thread pool \AgdaDatatype{ThreadPool}\AgdaSpace\AgdaBound{G}  is indexed by the session context which is split among all commands (cf.\
Figure~\ref{fig:resource-splitting-threads}). The empty 
thread pool \AgdaInductiveConstructor{tnil} holds no resources as indicated by
\AgdaDatatype{Inactive}\AgdaSpace\AgdaBound{G}. A non-empty thread
pool \AgdaInductiveConstructor{tcons} splits the resources between the
current thread and the rest of the thread pool.

Figure~\ref{fig:types-for-the-scheduler} contains type definitions for
the scheduler. Its core is a \AgdaFunction{step} function that attempts a single reduction on the thread pool and
the main scheduler executes steps in a round robin fashion. A step generates an \AgdaDatatype{Event}
that describes the reduction performed, if any. \AgdaInductiveConstructor{Terminated} means that no further
step can be taken because all threads have terminated. \AgdaInductiveConstructor{Restarted} means
that a \AgdaInductiveConstructor{Ready} thread has been
restarted. \AgdaInductiveConstructor{Stuck} means that no step is possible due to a deadlock
situation. We describe further events in the rest of this section, as they show up.

All events coming out of a \AgdaFunction{step} are accompanied by a thread pool.  The particular
session context after the step is obtained by context reduction as
specified in Figure~\ref{fig:context-reduction}. A context reduction
\AgdaOperator{\AgdaDatatype{\AgdaUnderscore{}⇒\AgdaUnderscore{}}}
either leaves the context unchanged (\AgdaInductiveConstructor{RedIdent}),
adds a new channel (\AgdaInductiveConstructor{RedNew}), or performs a
reduction step on an existing channel type
(\AgdaInductiveConstructor{RedInternal}) using the internal reduction relation
\AgdaOperator{\AgdaDatatype{\AgdaUnderscore{}⇒′\AgdaUnderscore{}}}.
This relation is structured analogously to de Bruijn indices. Either
the reduction takes place at a context entry below the top
(\AgdaInductiveConstructor{RedThere}),  the current channel gets
closed (\AgdaInductiveConstructor{RedEnd}), or the current channel
performs a transmission operation (\AgdaInductiveConstructor{RedTransmit}).
The \AgdaDatatype{NextPool} type encapsulates the event, the
reduction, and the resulting context. 

\begin{figure}[tp]
  \begin{flushleft}
    \ExecuteMetaData[Schedule3.tex]{step}
    \ExecuteMetaData[Schedule3.tex]{single-step}
  \end{flushleft}
  \caption{First steps}
  \label{fig:first-steps}
\end{figure}
Figure~\ref{fig:first-steps} contains the type of \AgdaFunction{step} and the first few cases. The
function traverses the thread pool using a zipper represented by two lists of type
\AgdaDatatype{ThreadPool}. The focus of the zipper is the head of the
first list. The initial call is shown in \AgdaFunction{single-step}. It passes the
current thread pool as the first argument and an empty pool as the second. The \AgdaFunction{step}
function considers commands from the first list and either executes them or moves them to the second
list. 

In the first case for \AgdaFunction{step}, both lists are empty and the program
terminates normally with the identity context reduction \AgdaInductiveConstructor{RedIdent}.

In the second case, the first list is empty, but the second is not. This configuration indicates a
deadlock because the \AgdaFunction{step} function was not able to execute any command in the thread
pool.

The third case deals with the \AgdaInductiveConstructor{Ready}
command. It applies the continuation to the value from the command. Operationally,
it resumes executing the body of the next enclosing
\AgdaInductiveConstructor{letbind} and 
decompose it. The thread is moved to the end of the pool (\AgdaFunction{tsnoc}) and the two parts of
the pool are merged (\AgdaFunction{tappend} and
\AgdaFunction{treverse}).\footnote{Round robin scheduling would put
  the current thread at the very end. We simplified this matter to avoid further trickery with splittings.}

The fourth case deals with the \AgdaInductiveConstructor{Halt} command. It indicates the termination
of an individual thread, which is removed from the thread pool. Again, the two parts of the thread
pool are merged. 

Before carrying on with further cases of the \AgdaFunction{step} function, we take a look at the
top-level scheduler.
\ExecuteMetaData[Schedule3.tex]{schedule}
The scheduler repeatedly applies the \AgdaFunction{single-step} function to a thread pool. At this
point we cannot easily predict termination, so the scheduler runs on a provided supply of
\AgdaDatatype{Gas}. It either terminates by issuing an \AgdaInductiveConstructor{OutOfGas} if the
gas is depleted, \AgdaInductiveConstructor{Terminated} if \AgdaFunction{single-step} came to an
end, or \AgdaInductiveConstructor{Stuck} if a deadlock was detected.  

To load an expression, we construct a one-element thread pool
from decomposing it and pass it to the scheduler.
% (Appendix~\ref{sec:loading-machine}).
\ExecuteMetaData[Schedule3.tex]{start}

\subsection{Channel Creation}
\label{sec:channel-creation}

\begin{figure}[tp]
  \ExecuteMetaData[Schedule3.tex]{step-new}
  \caption{Stepping --- the \AgdaInductiveConstructor{New} command}
  \label{fig:step-new}
\end{figure}
The \AgdaInductiveConstructor{New} command is quite instructive (see
Figure~\ref{fig:step-new}).
Creating a channel means to extend the global session context
\AgdaBound{G} with a new entry. Moreover, in the current thread the new entry must be
\AgdaInductiveConstructor{just} (\AgdaFunction{s}, \AgdaInductiveConstructor{POSNEG}), but
it must be \AgdaInductiveConstructor{nothing} for everybody else.

This extension happens rather indirectly in the code. The telltales are the uses of the splitting
constructor \AgdaInductiveConstructor{ss-left}, which indicate that the user of the topmost resource
is the \AgdaInductiveConstructor{VPair} value in the ready thread. Inside the pair,
\AgdaInductiveConstructor{ss-posneg} indicates that the channel resource is split into its two ends,
which are used up in the \AgdaInductiveConstructor{VChan} values. Both refer to the topmost resource
as indicated by the \AgdaInductiveConstructor{here-pos} and \AgdaInductiveConstructor{here-neg}
indices. This information together with the type of the continuation \Akappa{} fully determines the
type of the topmost resource. 

Moreover, the rest of the thread pool in \AgdaBound{tp} and \AgdaBound{tp2} as well as the
continuation \Akappa{} are still typed with the previous context \emph{before} creating the new channel.
Hence, the scheduler must weaken every other command in the thread pool and the continuation by
prepending \AgdaInductiveConstructor{nothing}. Weakening a command consists of weakening its contents,
recursively: continuations, value environments, values, and resource splittings. This
weakening is straightforward because we chose to expose all values in a thread
in traversable data types. Weakening is implemented by the \AgdaFunction{lift} functions. 

\subsection{Close and Wait}
\label{sec:wait-clos-chann}

\begin{figure}[tp]
  \begin{flushleft}
    \ExecuteMetaData[Schedule3.tex]{step-wait}
  \end{flushleft}
  \caption{Stepping --- the \AgdaInductiveConstructor{Wait} command}
  \label{fig:step-wait}
\end{figure}
A waiting thread just stays in the thread pool as shown in Figure~\ref{fig:step-wait}. The extra
code implements one of the ubiquitous resource rotations.

\begin{figure}[tp]
  \begin{flushleft}
    \ExecuteMetaData[Schedule3.tex]{step-close}
  \end{flushleft}
  \caption{Stepping --- the \AgdaInductiveConstructor{Close} command}
  \label{fig:step-close}
\end{figure}
\begin{figure}[tp]
  \ExecuteMetaData[AlternativeMatch3.tex]{vcr-match-body}
  \caption{Implementation of \AgdaFunction{vcr-match}}
  \label{fig:implementation-vcr-match}
\end{figure}
To execute the \AgdaInductiveConstructor{Close}
command (Figure~\ref{fig:step-close}), the scheduler checks all other threads for a matching
\AgdaInductiveConstructor{Wait} command (with essentially the same typing up to the
session type \AgdaInductiveConstructor{SEnd?}). The key part of this tedious matching is to
compare the channel references of two channel values, which is implemented in the
\AgdaFunction{vcr-match} function.
\ExecuteMetaData[AlternativeMatch3.tex]{vcr-match}
The implementation is surprisingly straightforward (see Figure~\ref{fig:implementation-vcr-match}) and the proof obligations for the result can only be
fulfilled by two matching channel ends. First, the channel references must refer to opposite ends
of the same channel. Second, there has to be an internal context reduction step that leads to an
inactive context. This requirement can be satisfied because the context \AgdaBound{G} that we start
with only holds the resources for the two channel ends. Closing this channel sets this resource to
inactive, which is also apparent from the \AgdaInductiveConstructor{RedEnd} reduction in
Figure~\ref{fig:context-reduction}.

The enclosing function \AgdaFunction{matchWaitAndGo} is much more tricky to implement. It searches
the thread pool for the wait command and actually closes the channel.

\ExecuteMetaData[AlternativeMatch3.tex]{matchWaitAndGo}

The parameter \AgdaBound{G} is the session context of the thread pool. It is first split into the
resources needed for the \AgdaInductiveConstructor{Close} command (\AgdaBound{Gc}), the channel to
be closed and the continuation, and the remaining thread pool (\AgdaBound{Gtp}). The thread pool
itself is passed as a worklist of type 
\AgdaDatatype{ThreadPool}\AgdaSpace\AgdaBound{Gtpwl}. The parameter of type
\AgdaDatatype{ThreadPool}\AgdaSpace\AgdaBound{Gtpacc} is used in recursive
calls as an accumulating parameter to collect commands that do not match the
\AgdaInductiveConstructor{Close}. 

Most cases of \AgdaFunction{matchWaitAndGo} are boring as they deal with
all non-\AgdaInductiveConstructor{Wait} constructors in the same way. In each case, there is a minor
resource rotation to account for reorganization of the pool. We show
the case for the \AgdaInductiveConstructor{New} command as an example.
\ExecuteMetaData[AlternativeMatch3.tex]{matchWaitAndGo-new}

The interesting case for the \AgdaInductiveConstructor{Wait} command is too involved to discuss it
here. The main complication is due to resource rotations, the actual work is performed by
\AgdaFunction{vcr-match}. 

\subsection{Send and Receive}
\label{sec:send-receive}
% \begin{figure}[tp]
%   \begin{minipage}[t]{0.58\linewidth}
%     \ExecuteMetaData[Syntax3.tex]{Expr-send}    
%   \end{minipage}
%   \begin{minipage}[t]{0.38\linewidth}
%     \ExecuteMetaData[Syntax3.tex]{Expr-recv}    
%   \end{minipage}
%   \caption{Syntax for send and receive}
%   \label{fig:syntax-send-receive}
% \end{figure}
\begin{figure}[tp]
  \begin{flushleft}
    %\begin{minipage}[t]{0.5\linewidth}
      \ExecuteMetaData[Session3.tex]{Command-Send}
    %\end{minipage}
    %\begin{minipage}[t]{0.45\linewidth}
      \ExecuteMetaData[Session3.tex]{Command-Recv}
    %\end{minipage}
  \end{flushleft}
  \caption{Commands for sending and receiving}
  \label{fig:commands-send-receive}
\end{figure}

Sending and receiving gives rise to two new commands for the scheduler,
\AgdaInductiveConstructor{Send} and \AgdaInductiveConstructor{Recv}
(Figure~\ref{fig:commands-send-receive}), which follow the same 
pattern as the commands discussed in Section~\ref{sec:commands}. 

\begin{figure}[tp]
  \ExecuteMetaData[AlternativeMatch3.tex]{vcr-match-2-sr}  
  \caption{Matching channel ends for sending and receiving}
  \label{fig:match-send-receive}
\end{figure}
The scheduler processes \AgdaInductiveConstructor{Send} and
\AgdaInductiveConstructor{Recv} analogously to \AgdaInductiveConstructor{Close} and
\AgdaInductiveConstructor{Wait}. A \AgdaInductiveConstructor{Send} command triggers a
search for a matching \AgdaInductiveConstructor{Recv} among the other threads. The core of
this search is again the matching of the channel ends implemented in
\AgdaFunction{vcr-match-2-sr}, but its typing (in
Figure~\ref{fig:match-send-receive}) is more involved. 

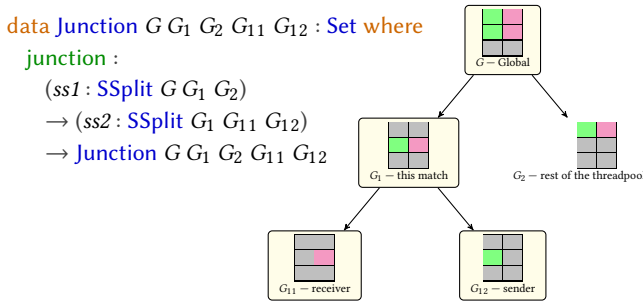
\begin{figure}[tp]
  \begin{flushleft}
    \begin{subfigure}[t]{0.4\columnwidth}
      \ExecuteMetaData[Channel3.tex]{SSplit2}
    \end{subfigure}
    % \hfill
    \begin{subfigure}[t]{0.1\columnwidth}
      \begin{center}
        \scalebox{.5}{\input{pic-resource-splitting-match}}
      \end{center}
  \end{subfigure}
    \end{flushleft}
    \caption{Splitting and changing the resource state
%       :\\
%       \AgdaDatatype{SSplit2}\AgdaSpace{}%
% \AgdaBound{G}\AgdaSpace{}%
% \AgdaBound{G₁}\AgdaSpace{}%
% \AgdaBound{G₂}\AgdaSpace{}%
% \AgdaBound{G₁₁}\AgdaSpace{}%
% \AgdaBound{G₁₂}.
}
\label{fig:split-change-resource}
\end{figure}
The main issue in this match is the transfer of resources between the
two threads. Three parties participate in this transfer, the receiving thread, the
sending thread, and the rest of the threadpool. The sender
and receiver may exchange resources as channels may be transferred
over channels, whereas the resources of the rest of the threadpool
remain unaffected. To implement this transfer in terms of splitting
trees, we rearrange the tree according to the type 
\AgdaDatatype{Junction} (Figure~\ref{fig:split-change-resource}). 
This single type (or its mirror image) with its two nested splittings is sufficient to manage all
resource transfers in a binary session type system because there are
never more than three parties involved, two active ones and one inactive. 

In the reduction step the receiving end
as well as the sending channel end change their type. Hence, all contexts on the path
up to the global context (yellow boxes in Figure~\ref{fig:split-change-resource}) need to make a
reduction step with relation $\Rightarrow'$ as indicated by the existentials in the type of
\AgdaFunction{vcr-match-2-sr}. The last two components of the output
tuple document this reduction. Furthermore, the function must 
prove that \AgdaBound{G₂}, which contains all remaining resources of
the system, remains unaffected by this update. The equality proofs are again
instrumental to demonstrate that the \AgdaFunction{vcr-match-2-sr}
function found the two ends of the same channel.

\begin{figure*}[tp]
  \begin{flushleft}
  \begin{subfigure}[t]{0.46\linewidth}
    \scalebox{.5}{\input{pic-resource-transfer}}
    \caption{Before}
    \label{fig:resource-transfer-before}
  \end{subfigure}
  \begin{minipage}[t]{1em}
    \raisebox{15ex}{$\Rightarrow$}
  \end{minipage}
  \begin{subfigure}[t]{0.48\linewidth}
    \scalebox{.5}{\input{pic-resource-transfer-after}}
    \caption{After}
    \label{fig:resource-transfer-after}
  \end{subfigure}
\end{flushleft}
\caption[Resource transfer]{Resource transfer. Updated splittings are highlighted by
    yellow boxes. An arrow indicates the transferred resource at the
    thread level. The actual resource is highlighted with a slightly
    darker box. Updated session types are marked with "x".}
  \label{fig:resource-transfer}
\end{figure*}
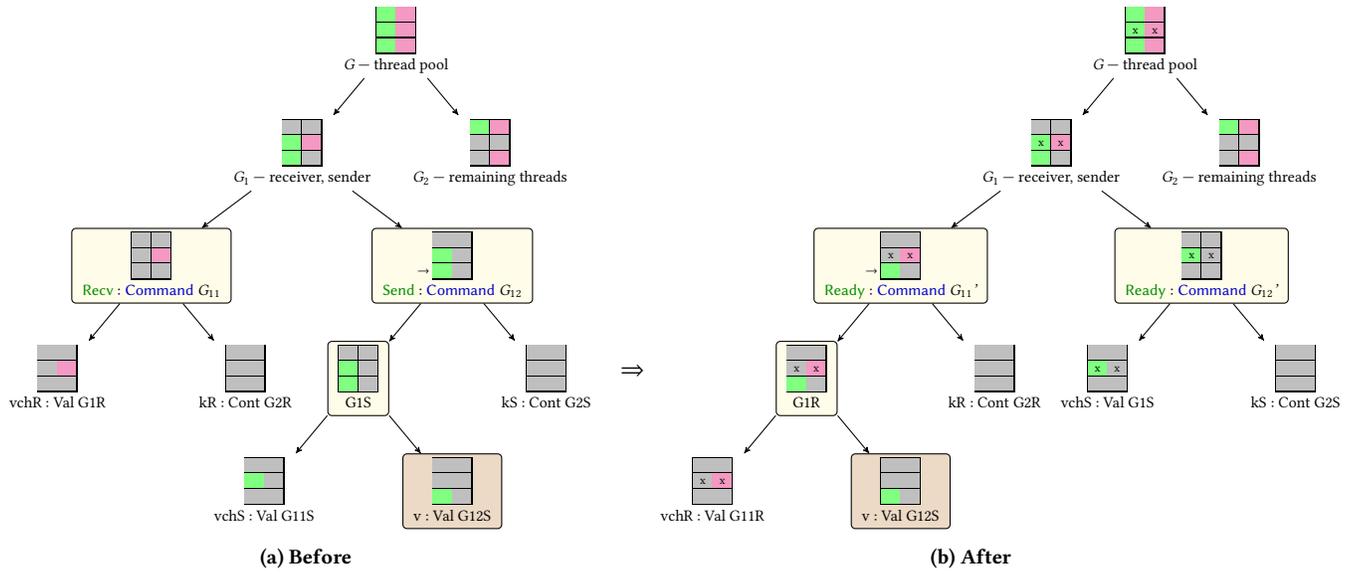
The scheduler actually transfers the resources from
sending thread ($G_{12}$) to
receiving thread ($G_{11}$). Figure~\ref{fig:resource-transfer} illustrates the
transfer with an example. 
Figure~\ref{fig:resource-transfer-before} (left) shows the splitting tree
before the communication and Figure~\ref{fig:resource-transfer-after} (right) illustrates the
splitting immediately after.  Before the communication the sender holds two resources: the
sending end of a channel \texttt{vchS} and the value \texttt{v} that
should be transmitted (in the brown box); the value \texttt{v} is
itself a channel end. The receiver holds just one resource: the
receiving end \texttt{vchR} corresponding to \texttt{vchS}. 
Hence, the transferred resource is a single
channel end \texttt{v} in this example, but in general it could be an arbitrary
chunk of resources.

After the communication (Figure~\ref{fig:resource-transfer-after}),
the receiver $G_{11}'$ holds two resources in the pair indexed with
\texttt{G1R}: the channel end \texttt{vchR} and 
the received value \texttt{v}. The sender $G_{12}'$ only retains the
channel end \texttt{vchS}, but ``lost'' the resource in \texttt{v}.
All context entries marked with ``XX'' change to account for the
context reduction that models the send/receive operation.

The commands in both threads 
change to the command \AgdaInductiveConstructor{Ready}, which holds
a pair of channel end and received value and the continuation after
receiving on the left and
the other channel end and the continuation after sending on the right.

The code for implementing the resource rotation shown in
Figure~\ref{fig:resource-transfer} is tedious and therefore omitted
from the paper. The definition of
type \AgdaDatatype{Junction}, which contains two nested splits arranged
as shown in Figure~\ref{fig:split-change-resource}, embodies an
important insight. It corresponds to
the tip of the resource tree in Figure~\ref{fig:resource-transfer} and
the insight is that this tip must be transformed as a whole.

\section{The Full Picture}
\label{sec:full-picture}

Building on the established basis of MicroSession, we discuss the addition of the
remaining ingredients of session typing, which are present in the calculi
of \CITET{Gay and Hole}{GayHole2005} and/or \CITET{Gay and Vasconcelos}{GayVasconcelos2010-jfp}. 
In particular, we consider
\begin{inparaenum}
%\item Sending and receiving values including channel-typed values (higher-order sessions).
\item internal and external choice,
\item functions and recursion, % (see appendix Sec.~\ref{sec:functions-recursion}),
\item recursive sessions,
\item subtyping,
\item structured gas,
\item and asynchrony.
\end{inparaenum}

\subsection{Internal and External Choice}
\label{sec:intern-extern-choice}

Internal and external choice require an additional session type constructors.
\ExecuteMetaData[Extension3.tex]{SChoice}
This constructor represents $m$-ary choice where \AgdaBound{m} is the number of
alternatives and \AgdaBound{cases} is a function from 
\AgdaDatatype{Fin}\AgdaSpace\AgdaBound{m} to \AgdaDatatype{STy}, where
\AgdaDatatype{Fin}\AgdaSpace\AgdaBound{m} is the finite type of
numbers $\{0,\dots,m-1\}$.\footnote{\AgdaDatatype{Data.Fin} from the
  Agda standard library.}

Expressions extend in the obvious way.
\ExecuteMetaData[Extension3.tex]{SChoice-syntax}
The expression
\AgdaInductiveConstructor{nselect}\AgdaSpace\AgdaBound{lab}\AgdaSpace\AgdaBound{ch}
corresponds to an internal choice of the alternative \AgdaBound{lab}
on channel \AgdaBound{ch}. The result is a channel of the type
selected by the \AgdaBound{lab}.

The expression \AgdaInductiveConstructor{nbranch}\AgdaSpace\AgdaBound{sp}\AgdaSpace\AgdaBound{ch}\AgdaSpace\AgdaBound{cases}
announces an external choice on channel \AgdaBound{ch}. 
The \AgdaBound{cases} argument holds a function that maps the alternative
number \AgdaBound{lab} to an expression that is abstracted over the corresponding
continuation channel of type \AgdaBound{(alt lab)}.

The corresponding new commands follow the familiar pattern.
\ExecuteMetaData[Extension3.tex]{SChoice-commands}
Resources are split between the channel value
\AgdaBound{vch}, on which the choice is made, and the continuation. For
internal choice (\AgdaInductiveConstructor{NSelect}), there is
only one continuation, whereas external choice
(\AgdaInductiveConstructor{NBranch}) selects its continuation
according to the label received.

The \AgdaFunction{step} function leaves the passive
\AgdaInductiveConstructor{NBranch} command in the thread pool. The
\AgdaInductiveConstructor{NSelect} command invokes a function
analogous to \AgdaFunction{matchWaitAndGo}, which in turn
relies on a channel match similar to \AgdaFunction{vcr-match}. As
there are no resources to transfer between the threads, the resource
rotation is less involved than in Section~\ref{sec:send-receive}. 

Alternatively, we can follow
\citet{DBLP:conf/esop/LindleyM15} or \citet{DBLP:conf/esop/Padovani17}
who encode choice using sum types. However, Lindley
and Morris create a new channel at every choice,
which would change our invariants about resource usage, whereas 
Padovani requires clever retyping of the channel at the
receiving end.

\subsection{Recursive Sessions}
\label{sec:recursive-sessions}
To extend our model to recursive sessions, we revise the
type structure following \citet{DBLP:journals/jfp/AbelAS17}.
% In the appendix (Figure~\ref{fig:recursive-sessions}), we show how
Session types are modeled as the greatest fixed point of a functor
\AgdaDatatype{STF}, which models the different session type
constructors. 
The greatest fixed point is obtained with a coinductive record type
(\AgdaDatatype{SType}). The channel type
\AgdaInductiveConstructor{TChan} refers to the unraveled
session type \AgdaDatatype{STF}\AgdaSpace\AgdaDatatype{SType},
on which we can pattern match (unlike the
type \AgdaDatatype{SType}, the values of which need to be forced).
This change causes only minor
modifications of the code.
\begin{itemize}
\item We define a coinductive type equivalence relation and
  prove its reflexivity, symmetry, and transitivity. Definitions
  follow the same pattern as for \AgdaDatatype{STF} and \AgdaDatatype{SType}.
  Proofs are tedious, but straightforward. Uses of propositional equality on \AgdaDatatype{Type}
  are replaced by explicit use of the equivalence relation.
\item We provide a coinductive definition of duality along
  with a coinductive proof that duality is involutory. 
\item The types of expressions that manipulate session types 
  force their session typed arguments, as in the type of
  \AgdaInductiveConstructor{new}: 
\ExecuteMetaData[Extension3.tex]{rectypes-new}  
\end{itemize}

\subsection{Subtyping}
\label{sec:subtyping}

Subtyping of session types arises from the choice types. Our notion of
subtyping derives from Liskov's substitution principle as in GV
\cite{GayVasconcelos2010-jfp}. A channel with an internal choice can
be subsumed to a choice with fewer alternatives (the
supertype). Dually, a channel with an external choice can be subsumed
to a choice with more alternatives, as it is always ok to accept more
alternatives than needed.

In our modeling, a selection (internal choice) on a
\AgdaDatatype{Fin}\AgdaSpace\AgdaBound{m} can be subsumed to any
\AgdaDatatype{Fin}\AgdaSpace$m'$ with $m'\le m$.
Analogously, a branch (external choice) on a
\AgdaDatatype{Fin}\AgdaSpace\AgdaBound{m} can be subsumed to any
\AgdaDatatype{Fin}\AgdaSpace$m'$ with $m\le m'$.
In connection with recursive session types, the definition of subtyping is also
coinductive. It uses the same style as the coinductive definition of
session types.
%(see appendix Fig.~\ref{fig:subtyping-recursive-session-types}).

The remaining changes in the framework are
straightforward. Expressions are extended with an explicit subsumption
constructor that implements the type subsumption rule. At run time,
subsumption is implemented by a coercion that adjusts the type of
channel values and expressions inside of continuations.

\subsection{Adequacy}
\label{sec:adequacy}

We investigated some metatheoretical properties of our model. As
we implement the lambda calculus reductions indirectly via the
stuttering CEK machine, an obvious question to ask is whether
execution is adequate with respect to standard lambda calculus
reductions.

For example, we investigate adequacy of beta reduction for functions
and pairs in all contexts. In module \AgdaModule{Properties.StepBeta} we show that
$\mathtt{let}\ f = (\lambda x.e)\ \mathtt{in}\ \mathtt{let}\ z = (f\
y)\ \mathtt{in}\ E$ is equal to $\mathtt{let}\ z = e[y/x]\
\mathtt{in}\ E$.
In module \AgdaModule{Properties.StepPair} we show that
$\mathtt{let}\ p= (a,b)\ \mathtt{in}\ \mathtt{let}\ (x,y) = p\
\mathtt{in}\ E$ is equal to $E[a,b/x,y]$. In both cases, equality is
formalized as identity after a fixed number of steps as in
\AgdaFunction{restart}\AgdaSpace(\AgdaFunction{restart}\AgdaSpace\AgdaBound{lhs})\AgdaSpace\AgdaSymbol{≡}\AgdaSpace\AgdaBound{rhs}. Here
\AgdaBound{lhs} is the machine state loaded with the term before
reduction and \AgdaBound{rhs} is the same after reduction. Function
\AgdaFunction{restart} performs one step of the CEK machine.
These modules are available in the GitHub repository.

There are analogous results about the execution of \AgdaInductiveConstructor{close}/\AgdaInductiveConstructor{wait}, \AgdaInductiveConstructor{fork}, and
\AgdaInductiveConstructor{new}. However, these statements and their
proofs refer to process calculus reductions and they are restricted to
very simple contexts. The proofs for general 
contexts get very complex and we have not been able to complete them.

% \section{Extensions}
% \label{sec:extensions}

% This section discusses extensions that might be desirable, but
% have not been implemented, yet. 

\subsection{Structured Gas}
\label{sec:structured-gas}

The present scheduler relies on a \AgdaDatatype{Gas} type that is isomorphic to
the natural numbers. As the scheduler is a function, the
execution only covers one particular schedule. To obtain arbitrary
interleavings, we may adopt \textbf{structured gas}, which is isomorphic
to a list of numbers. Whenever the scheduler is invoked with a gas element
\AgdaBound{n}, it rotates the thread pool by $n$ positions before
trying to make a step. As the interpreter
works with any structured gas value, any schedule can be explored in this way.

\subsection{Asynchrony}
\label{sec:asynchrony}

The presented framework models synchronous communication and it is
obvious to ask for a model of asynchronous communication.
The standard formalization in the literature
\cite{GayVasconcelos2010-jfp}  adds extra queue processes that
keep sent messages until their receiver is ready to accept them.
% In this setting, the send operation moves the value to the queue and
% continues immediately.
In contrast, we encode an
asynchronous channel of type $s$ as a \emph{promise} for such a
channel. A promise for $s$ is a single-shot synchronous channel from which we can
receive a synchronous payload channel of type $s$.
\ExecuteMetaData[Typing3.tex]{async}
Our scheme represents one asynchronous channel by three synchronous
channels: the payload channel and one promise channel at each end. The
promise channel is replaced by each operation so that there is a chain
of promises at each end which simulates an unbounded buffer.
We achieve asynchronous operation by performing all potentially
blocking operations on promise and payload channels in a separate
thread.

Asynchronous sending creates a new promise channel (with ends
\texttt{receiver} and
\texttt{sender}), forks a process to 
perform the send, and immediately returns the receiving end of the
promise channel. The forked process synchronously receives the actual
channel \texttt{schan} from the previous promise \texttt{achan}, sends the value, and finally sends
the depleted channel to the new promise. Here is the pseudocode:
\begin{lstlisting}
asend achan value =
  let (receiver, sender) = new in
  fork (let (schan, achan') = recv achan in
        let schan' = send schan value in
        let sender' = send sender schan' in
        let _ = close achan' in       
        wait sender');
  receiver
\end{lstlisting}

Asynchronous receive blocks on the promise \texttt{achan} and then on the received
channel \texttt{schan}. It creates a new promise, which needs to be filled in a
separate thread.
\begin{lstlisting}
arecv achan =
  let (schan, achan') = recv achan in
  let (value, schan') = recv schan in
  let (receiver, sender) = new in
  fork (let sender' = send sender schan' in
        let _ = close achan' in 
        close sender');
  (value, receiver)
\end{lstlisting}

Finally, we exhibit the signatures of the basic asynchronous
operations. The implementation consists in each case of an expression
of the appropriate type. They are straightforward to construct from
the pseudocode given above.

\ExecuteMetaData[Async3.tex]{async-anew}
\ExecuteMetaData[Async3.tex]{async-asend}
\ExecuteMetaData[Async3.tex]{async-arecv}

This definition applies to the simple, non-recursive case, but it
scales to the full system with recursive types and subtyping.

% \subsection{Multiparty Session Types}
% \label{sec:multiparty}

% Honda and others extended session types to more than two parties
% \cite{HondaYoshidaCarbone2008}. We believe that our model can be
% extended to cover synchronous multiparty session types by adopting a
% suitable multi-headed representation for the session context. 

\section{Related Work}
\label{sec:related}

\citet{DBLP:journals/mscs/GotoJJPR16} provide a Coq proof for subject reduction and
safety properties of a variant of pi calculus with session types,
polymorphism, and name matching. Their formalization
differs substantially from ours. Processes have an untyped, locally-nameless
representation. Congruence of the process language is modeled
explicitly. Reduction is modeled as a relation on untyped terms. Their
session type language includes neither recursion nor
subtyping. % To support polymorphism,
Duality is expressed as a type
constructor rather than as a function on types. Linearity is maintained using a partitioning relation
analogous to our splitting of type environments in the syntax.

\citet{DBLP:journals/corr/OrchardY16} establish a type-preserving translation
between imperative processes with effect types and $\pi$-calculus with
session types and prove that the translation is sound with respect to
an equational theory. That work comes with an intrinsically typed Agda formalization of
(part of) the syntax and the translation. Recursive types are not
formalized and subtyping is very limited. No attempt is made at
formalizing the semantics or the equational theory.

\citet{DBLP:journals/mscs/PereraC18} formalize a proof of a diamond
lemma for concurrent execution in $\pi$-calculus in Agda. Their
calculus is untyped, the formalization relies on de Bruijn encoding
for binders and encodes the semantics using an inductive relation.

The superstructure of our proof is inspired by the soundness proofs by
\citet{GayVasconcelos2010-jfp} and \citet{DBLP:conf/esop/Padovani17}.
Polarities were introduced by \CITET{Pierce and
  coworkers}{DBLP:conf/lics/PierceS93} to describe the direction of
communication on a channel. \CITET{Gay and Hole}{GayHole2005} use them
in the context of session types 
% for the pi calculus
with a slightly
different meaning, to distinguish the two ends of a communication
channel.  Our use of polarities corresponds to the latter work.
% Connection to polarities (Gay and Hole \cite{GayHole2005}). 

The use of indexing to tame potential nontermination can be traced back to
\citet{DBLP:journals/toplas/AppelM01}. It has been rejuvenated in the
context of big-step semantics by
\citet{siek13:_type_safet_three_easy_lemmas} and
\citet{DBLP:conf/esop/OwensMKT16}, in proving type soundness with
definitional interpreters \cite{DBLP:conf/popl/AminR17}, and in
definitional interpretation for imperative languages
\cite{DBLP:journals/pacmpl/PoulsenRTKV18}. 
As these interpreters are universally quantified over gas,
the respective properties hold for all terminating and nonterminating
computations. 
\citet{DBLP:conf/sbmf/Wadler18} uses the same device to
execute small-step semantics in Agda.

Alternatively, we could make use of the partiality monad
\cite{DBLP:conf/icfp/Danielsson12,DBLP:journals/lmcs/Capretta05}, but
that approach is better suited to big-step, interpretive settings where there are
no intermediate results. A gas-driven semantics really shines in a
small-step setting as it enables inspecting all intermediate states of
a computation.

% \begin{itemize}
% \item Coinductive Big-Step Semantics for Concurrency \cite{DBLP:journals/corr/Uustalu13}
% \item Hoare Logic for the Coinductive Trace-Based Big-Step Semantics
%   \cite{DBLP:conf/esop/NakataU10} 
% \item Resumptions etc \cite{DBLP:journals/corr/abs-1008-2112}
% \end{itemize}

% Our use of commands is reminiscent to free monads and work on effect
% handlers. 

% Comparison of typed typed with small-step and big-step approaches.

% Relation to Vasco's paper with Giunti \cite{DBLP:journals/mscs/GiuntiV16}

\section{Conclusions}
\label{sec:conclusions}

We claim type soundness for a functional session type calculus.  Type
preservation holds because all objects manipulated by the semantics
are intrisically typed, from syntax to values, and this property is
preserved throughout. Linearity is preserved by maintaining the
splitting of variable environments and resources.
We can be sure that there is no loophole because all Agda programs are terminating.

Progress is embodied in the definition of the \AgdaFunction{step} function
(Section~\ref{sec:scheduler}). By design, this function returns
\AgdaInductiveConstructor{Stuck}, if it cannot make progress. That is,
there is no command in the thread pool that can execute
immediately and all remaining communication commands do not have a
corresponding partner at the other end of the channel. Stating such a
relative progress result feels like a futile
exercise because its definition and proof
mirror exactly the definitions of \AgdaFunction{step}, the
\AgdaFunction{match*AndGo} functions, and the
\AgdaFunction{vcr-match*} functions. 

What was hard about constructing this model?
The key issues in the construction of this framework are
\begin{inparaenum}
\item 
  the management of resources by maintaining the splitting tree,
\item
  the identification of the right notion of \AgdaDatatype{Command} and
  its resource assignment,
\item
  the realization that a sleeping thread cannot be affected by
  executing another command: by linearity, its own resources cannot change and any other
  resources remain unavailable; new resources only enter through the
  continuation; weakening can be performed transparently, if the other
  command creates a new channel.
\end{inparaenum}

Our framework is a good starting point for further
investigation towards modeling multiparty session
types. It will be interesting to see if 
recent developments, e.g.,  in quantative type theory
\cite{DBLP:conf/lics/Atkey18}, can be used to simplify the handling of linearity.
It would also be fruitful to further explore the connection of
splitting with separation logic with the goal of obtaining cleaner,
more modular proofs. Some authors propose modeling session
types with separation logic \cite{DBLP:conf/aplas/CosteaCQC18}, but
further investigation is needed.

%% Acknowledgments
\begin{acks}                            %% acks environment is optional
                                        %% contents suppressed with 'anonymous'
  Thanks to the reviewers for their helpful comments.

  %% Commands \grantsponsor{<sponsorID>}{<name>}{<url>} and
  %% \grantnum[<url>]{<sponsorID>}{<number>} should be used to
  %% acknowledge financial support and will be used by metadata
  %% extraction tools.
  % This material is based upon work supported by the
  % \grantsponsor{GS100000001}{National Science
  %   Foundation}{http://dx.doi.org/10.13039/100000001} under Grant
  % No.~\grantnum{GS100000001}{nnnnnnn} and Grant
  % No.~\grantnum{GS100000001}{mmmmmmm}.  Any opinions, findings, and
  % conclusions or recommendations expressed in this material are those
  % of the author and do not necessarily reflect the views of the
  % National Science Foundation.
\end{acks}

\clearpage
%% Bibliography
%\bibliography{bibfile}
\bibliography{abbrv,local,papers,books,collections,misc,theses}

%\end{document}

%% Appendix
\clearpage
\appendix
\section{Supplementary Material}
\subsection{Processes}
\label{sec:processes}
\begin{figure}[h]
    \ExecuteMetaData[ProcessSyntax3.tex]{proc-data}
    \caption{Syntax of processes}
    \label{fig:syntax-processes}
\end{figure}

The GV calculus adds a separate language of processes
on top of expressions as defined in
Fig.~\ref{fig:syntax-processes}. Processes are also intrinsically typed
and hence indexed by a typing context \APhi. This context only
contains bindings for channel endpoints.
A process $P$ is either an expression process
(\AgdaInductiveConstructor{exp}), a parallel execution of processes
(\AgdaInductiveConstructor{par}) which splits the environment for its
two subprocesses, or a
restriction (\AgdaInductiveConstructor{res}) with introduces a channel
of session type $s$ scoped over the process $P$. For typing a
restriction we use the technique of \emph{polarities} 
adapted from \CITET{Gay and Hole}{GayHole2005}.  Restriction introduces a single bidirectional channel,
which materializes in the type environment as two channel ends adorned
with polarities, as in $c^+$ and $c^-$. The positive end has the
declared session type of the channel, whereas the negative end carries the dual type.

In standard treatments, process syntax is subject to the usual structural congruence rules: a unit
expression process $\PExp{()}$ can be discarded, the parallel operator is commutative and associative,
restrictions may be swapped and commute with the parallel operator subject to alpha
renaming of the bound channel.

% \subsection{Loading the Machine}
% \label{sec:loading-machine}

\subsection{Resource Splitting}
\label{sec:resource-splitting}
  \begin{figure}[tp]
    \centering \scalebox{0.5}{\input{pic-resource-splitting-values}}
    \caption{Splitting among values}
    \label{fig:resource-splitting-values}
  \end{figure}
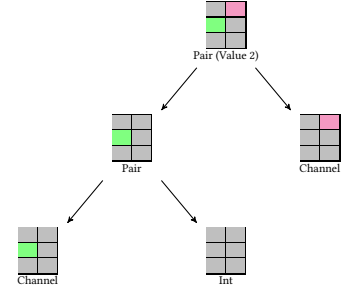
 The splitting structure of value environments is analogous to thread
pools. For values, we observe that primitive values hold no resources, a pair holds the
disjoint union of the resources of its components, and a channel value must hold exactly
one resource identifying the channel and its polarity as shown in
Figure~\ref{fig:resource-splitting-values}.

As an example lemma, we prove that splitting is symmetric. This
property is needed in Section~\ref{sec:scheduler}.
\ExecuteMetaData[Global0.tex]{ssplit-sym}

\subsection{Functions and Recursion}
\label{sec:functions-recursion}
\begin{figure}[tp]
  \begin{flushleft}
\ExecuteMetaData[Extension3.tex]{Function-syntax}  
\ExecuteMetaData[Extension3.tex]{Function-syntax-app}  
  \end{flushleft}
%   \begin{subfigure}[t]{0.45\linewidth}
% \ExecuteMetaData[Extension3.tex]{Function-syntax}  
%   \end{subfigure}
%   \qquad
%   \begin{subfigure}[t]{0.45\linewidth}
% \ExecuteMetaData[Extension3.tex]{Function-syntax-app}  
%   \end{subfigure}
  \caption{Expressions for function types}
  \label{fig:expressions-function-types}
\end{figure}

The extension with functions and recursion is standard, except for
dealing with linearity. 
Figure~\ref{fig:expressions-function-types} provides two forms for defining functions,
one for linear-use functions
(\AgdaInductiveConstructor{llambda}) that can close over any kind of variables
and another for recursive functions (\AgdaInductiveConstructor{rec}) that can only contain unrestricted free
variables. 

A linear-use function \AgdaInductiveConstructor{llambda} places no restrictions on the
environment. The function type constructor 
\AgdaDatatype{TFun} takes an extra argument
\AgdaInductiveConstructor{LL} that enforces linear use of the function.

A recursive function \AgdaInductiveConstructor{rec} requires that the entire environment is
unrestricted and constructs an unrestricted function indicated by the
\AgdaInductiveConstructor{UU} tag on the function type constructor.

Function application \AgdaInductiveConstructor{app} is parametric in
the linearity indicator of the function type.

Evaluation of a recursive function unrolls it into a function value
and puts it into a \AgdaInductiveConstructor{Ready}
command. Function application extends the current continuation with
the function body and creates a \AgdaInductiveConstructor{Ready}
command with the function's argument. This way, dealing with
potentially nonterminating functions is left to the gas-driven
scheduler loop.

\begin{figure*}[tp]
  \begin{subfigure}[t]{0.55\linewidth}
\ExecuteMetaData[Extension3.tex]{rectypes}    
  \end{subfigure}
  \quad
  \begin{subfigure}[t]{0.4\linewidth}
\ExecuteMetaData[Extension3.tex]{rectypes-1}    
  \end{subfigure}
  \caption{Recursive sessions}
  \label{fig:recursive-sessions}
\end{figure*}

\begin{figure*}
  \ExecuteMetaData[Extension3.tex]{rectypes-sub}
  \caption{Subtyping of recursive session types}
  \label{fig:subtyping-recursive-session-types}
\end{figure*}

\subsection{Subtyping}
\label{sec:subtyping-appendix}

Fig.~\ref{fig:subtyping-recursive-session-types} specifies the
subtyping relation for recursive sessions. 
It is a coinductive definition following the same pattern as for the
session types.
All nontrivial subtyping is generated by the two constructors / rules
\AgdaInductiveConstructor{sub-sint} and
\AgdaInductiveConstructor{sub-sext} corresponding to internal and
external choice. 
For the internal choice, the supertype has fewer alternatives $m'\le
m$. The premises of the rule establish the subtype relation just for
the first $m'$ alternatives. The injection function
\AgdaFunction{inject$\le$} from the standard library adjusts the type
of $i$ from 
\AgdaDatatype{Fin}\AgdaSpace\AgdaBound{$m'$} to 
\AgdaDatatype{Fin}\AgdaSpace\AgdaBound{m} given a proof that $m'\le
m$. External choice is handled dually.

% \subsection{Asynchrony Implementation}
% \label{sec:asynchrony-implementation}

\end{document}

%% file: macros.tex
\newcommand\CITET[2]{\citet{#2}}
\newcommand\Chan[3][\phantom{x}]{%
  \begin{tabular}{|c|c|}
    \hline
    \cellcolor{#2}#1&\cellcolor{#3}#1
    \\\hline
  \end{tabular}%
}
\newcommand\ChanNothing{\Chan{gray!50}{gray!50}}
\newcommand\ChanPos{\Chan{green!50}{gray!50}}
\newcommand\ChanNeg{\Chan{gray!50}{magenta!50}}
\newcommand\ChanPosNeg{\Chan{green!50}{magenta!50}}

\newcommand\ChanPosUpdated{\Chan[x]{green!50}{gray!50}}
\newcommand\ChanNegUpdated{\Chan[x]{gray!50}{magenta!50}}
\newcommand\ChanPosNegUpdated{\Chan[x]{green!50}{magenta!50}}

\newcommand\RArrow{\makebox[0pt][r]{$\rightarrow$ }}

\newcommand\Hole\Box
\newcommand\stepsto\longrightarrow

\newcommand\syntax[1]{\textup{\textsf{#1}}}

\newcommand\SEndS{\mathsf{end!}}

\newcommand\PExp[1]{\langle#1\rangle}

\newcommand\APhi{\AgdaBound{Φ}}

\newcommand\Arho{\AgdaBound{ϱ}}
\newcommand\Akappa{\AgdaBound{κ}}

%% file: fig-microsession-example.tex
\begin{figure}[tp]
  \begin{subfigure}[t]{0.45\linewidth}
    \begin{gather*}
      \begin{array}{l}
        \mathtt{let}\ x\ \mathtt{=}\ (\mathtt{new}\ \SEndS)\
        \mathtt{in}\ \\
        \mathtt{let}\ (c_1, c_2)\ \mathtt{=}\ x\ \mathtt{in}\ \\
        \mathtt{let}\ u\ \mathtt{=}\ \mathtt{fork}\ (\mathtt{close}\ c_1)\
        \mathtt{in}\ \\
        \mathtt{wait}\ c_2
      \end{array}
    \end{gather*}
    \caption{Human readable}
    \label{fig:example-in-math}
  \end{subfigure}
  \begin{subfigure}[t]{0.45\linewidth}
    \ExecuteMetaData[Examples0.tex]{ex1block}
    \caption{Agda encoding}
    \label{fig:example-in-agda}
  \end{subfigure}
\caption{Example term}
\label{fig:micro-session-example}
\end{figure}

%% file: pic-resource-splitting-threads.tex
  \begin{tikzpicture}[->,>=stealth',auto,node distance=2.5cm,thick]
    \node (tproot) {
      \begin{tabular}{c}
        \ChanPosNeg\\\ChanPosNeg\\\ChanNothing
        \\Pool
      \end{tabular}
    } ;
    \node (tprootleft) [left of=tproot] {} ;
    \node (tprootright) [right of=tproot] {} ;
    \node (thread1) [below of=tprootleft] {
      \begin{tabular}{c}
        \ChanPos\\\ChanNothing\\\ChanNothing
        \\Thread 1
      \end{tabular}
    } ;
    \node (tpool1) [below of=tprootright] {
      \begin{tabular}{c}
        \ChanNeg\\\ChanPosNeg\\\ChanNothing
      \end{tabular}
    } ;
    \node (tpool1left) [left of=tpool1] {} ;
    \node (tpool1right) [right of=tpool1] {};
    \node (thread2) [below of=tpool1left] {
      \begin{tabular}{c}
        \ChanNeg\\\ChanPos\\\ChanNothing
        \\Thread 2
      \end{tabular}
    } ;
    \node (tpool2) [below of=tpool1right] {
      \begin{tabular}{c}
        \ChanNothing\\\ChanNeg\\\ChanNothing
      \end{tabular}
    };
    \node (tpool2left) [left of=tpool2] {} ;
    \node (tpool2right) [right of=tpool2] {};
    \node (thread3) [below of=tpool2left] {
      \begin{tabular}{c}
        \ChanNothing\\\ChanNeg\\\ChanNothing
        \\Thread 3
      \end{tabular}
    };
    \node (tpool3) [below of=tpool2right] {
      \begin{tabular}{c}
        \ChanNothing\\\ChanNothing\\\ChanNothing
        \\Empty
      \end{tabular}
    };
    \path  (tproot) edge (thread1) ;
    \path  (tproot) edge (tpool1) ;
    \path  (tpool1) edge (thread2) ;
    \path  (tpool1) edge (tpool2) ;
    \path  (tpool2) edge (thread3) ;
    \path  (tpool2) edge (tpool3) ;
  \end{tikzpicture}

%%% Local Variables: 
%%% mode: latex
%%% TeX-master: "popl19-defint"
%%% End: 
  

%% file: pic-ssplit-compose-left0.tex
\begin{tikzpicture}[->,>=stealth',
  level/.style={sibling distance = 2cm, level distance = 1cm},
  thick,
  affected/.style={rectangle, draw, fill=yellow!10,rounded corners=.8ex}
  ]
      \node {%
        \AgdaBound{G}
      }
      child {
        node {%
          \AgdaBound{G₁}
        }
          child {
            node {%
            \AgdaBound{G₁₁}
            }
          }
          child {
            node {%
            \AgdaBound{G₁₂}
            }
          }
      }
      child {
          node {%
            \AgdaBound{G₂}
          }
        }
      ;
    \end{tikzpicture}

%%% Local Variables: 
%%% mode: latex
%%% TeX-master: "popl19-defint"
%%% End: 

%% file: pic-ssplit-compose-right0.tex
\begin{tikzpicture}[->,>=stealth',
  level/.style={sibling distance = 2cm, level distance = 1cm},
  thick,
  affected/.style={rectangle, draw, fill=yellow!10,rounded corners=.8ex}
  ]
      \node {%
        \AgdaBound{G}
      }
      child {
        node {%
          \AgdaBound{G₁₁}
        }
      }
      child {
        node [affected] {%
          \AgdaBound{Gi}
        }
        child {
          node {%
          \AgdaBound{G₁₂}
          }
        }
        child {
          node {%
            \AgdaBound{G₂}
          }
        }
      }
      ;
    \end{tikzpicture}

%%% Local Variables: 
%%% mode: latex
%%% TeX-master: "popl19-defint"
%%% End: 

%% file: pic-resource-splitting-match.tex
    \begin{tikzpicture}[->,>=stealth',level/.style={sibling distance = 5cm,
        level distance = 3cm},thick,
      affected/.style={rectangle, draw, fill=yellow!10,rounded corners=.8ex}
      ]
      \node [affected] {%
        \begin{tabular}{c}
          \ChanPosNeg\\\ChanPosNeg\\\ChanNothing
          \\\AgdaBound{G} --- Global
        \end{tabular}
      }
      child {
        node [affected] {%
          \begin{tabular}{c}
            \ChanNothing\\\ChanPosNeg\\\ChanNothing
            \\\AgdaBound{G₁} --- this match
          \end{tabular}
        }
        child {
          node [affected] {%
            \begin{tabular}{c}
              \ChanNothing\\\ChanNeg\\\ChanNothing
              \\\AgdaBound{G₁₁} --- receiver
            \end{tabular}
          }
        }
        child {
          node [affected] {%
            \begin{tabular}{c}
              \ChanNothing\\\ChanPos
              \\\ChanNothing
              \\\AgdaBound{G₁₂} --- sender
            \end{tabular}
          }
        }
      }
      child {
          node  {%
            \begin{tabular}{c}
              \ChanPosNeg\\\ChanNothing\\\ChanNothing
              \\\hspace{-3em}\AgdaBound{G₂} --- rest of the threadpool
            \end{tabular}
          }
      }
      ;
    \end{tikzpicture}

%%% Local Variables: 
%%% mode: latex
%%% TeX-master: "popl19-defint"
%%% End:

%% file: pic-resource-transfer.tex
\begin{tikzpicture}[->,>=stealth',
  level/.style={sibling distance = 5cm, level distance = 3cm},
  thick,
  affected/.style={rectangle, draw, fill=yellow!10,rounded corners=.8ex},
  transferred/.style={rectangle, draw, fill=brown!30,rounded corners=.8ex}
  ]
      \node {%
        \begin{tabular}{c}
          \ChanPosNeg\\\ChanPosNeg\\\ChanPosNeg
          \\[1.0ex]\LARGE\AgdaBound{G} --- thread pool
        \end{tabular}
      }
      child {
        node {%
          \begin{tabular}{c}
            \ChanNothing\\\ChanPosNeg\\\ChanPos
            \\[1.0ex]\LARGE\AgdaBound{G₁} --- receiver, sender
          \end{tabular}
        }
        child [sibling distance = 8cm] {
          node [affected] {%
            \begin{tabular}{c}
              \ChanNothing\\\ChanNeg\\\ChanNothing
              \\[1.0ex]\LARGE\AgdaInductiveConstructor{Recv} : \AgdaDatatype{Command}\AgdaSpace\AgdaBound{G₁₁}
            \end{tabular}
          }
          child {
            node {%
            \begin{tabular}{c}
              \ChanNothing\\\ChanNeg\\\ChanNothing
              \\[1.0ex]\LARGE vchR : Val G1R
            \end{tabular}
            }
          }
          child {
            node {%
            \begin{tabular}{c}
              \ChanNothing\\\ChanNothing\\\ChanNothing
              \\[1.0ex]\LARGE kR : Cont G2R
            \end{tabular}
            }
          }
        }
        child [sibling distance = 8cm] {
          node [affected] {%
            \begin{tabular}{c}
              \ChanNothing\\\ChanPos
              \\\RArrow\ChanPos
              \\[1.0ex]\LARGE\AgdaInductiveConstructor{Send} : \AgdaDatatype{Command}\AgdaSpace\AgdaBound{G₁₂}
            \end{tabular}
          }
          child {
            node [affected] {%
              \begin{tabular}{c}
                \ChanNothing\\\ChanPos
                \\\ChanPos
                \\[1.0ex]\LARGE G1S
              \end{tabular}
            }
            child {
              node {%
                \begin{tabular}{c}
                  \ChanNothing\\\ChanPos
                  \\\ChanNothing
                  \\[1.0ex]\LARGE vchS : Val G11S
                \end{tabular}
              }
            }
            child {
              node [transferred] {%
                \begin{tabular}{c}
                  \ChanNothing\\\ChanNothing
                  \\\ChanPos
                  \\[1.0ex]\LARGE v : Val G12S
                \end{tabular}
              }
            }
          }
          child {
            node {%
              \begin{tabular}{c}
                \ChanNothing\\\ChanNothing
                \\\ChanNothing
                \\[1.0ex]\LARGE kS : Cont G2S
              \end{tabular}
            }
          }
        }
      }
      child {
          node  {%
            \begin{tabular}{c}
              \ChanPosNeg\\\ChanNothing\\\ChanNeg
              \\[1.0ex]\LARGE\AgdaBound{G₂} --- remaining threads
            \end{tabular}
          }
      }
      ;
    \end{tikzpicture}

%%% Local Variables: 
%%% mode: latex
%%% TeX-master: "ppdp19-mechanized"
%%% End: 

%% file: pic-resource-transfer-after.tex
\begin{tikzpicture}[->,>=stealth',
  level/.style={sibling distance = 5cm, level distance = 3cm},
  thick,
  affected/.style={rectangle, draw, fill=yellow!10,rounded corners=.8ex},
  transferred/.style={rectangle, draw, fill=brown!30,rounded corners=.8ex}
  ]
      \node {%
        \begin{tabular}{c}
          \ChanPosNeg\\\ChanPosNegUpdated\\\ChanPosNeg
          \\[1.0ex]\LARGE \AgdaBound{G} --- thread pool
        \end{tabular}
      }
      child {
        node {%
          \begin{tabular}{c}
            \ChanNothing\\\ChanPosNegUpdated\\\ChanPos
            \\[1.0ex]\LARGE \AgdaBound{G₁} --- receiver, sender
          \end{tabular}
        }
        child [sibling distance = 8cm] {
          node [affected] {%
            \begin{tabular}{c}
              \ChanNothing\\\ChanNegUpdated\\
              \RArrow\ChanPos
              \\[1.0ex]\LARGE \AgdaInductiveConstructor{Ready} : \AgdaDatatype{Command}\AgdaSpace\AgdaBound{G₁₁'}
            \end{tabular}
          }
          %%%
          child {
            node [affected] {%
              \begin{tabular}{c}
                \ChanNothing\\\ChanNegUpdated
                \\\ChanPos
                \\[1.0ex]\LARGE G1R
              \end{tabular}
            }
            child {
              node {%
                \begin{tabular}{c}
                  \ChanNothing\\\ChanNegUpdated
                  \\\ChanNothing
                  \\[1.0ex]\LARGE vchR : Val G11R
                \end{tabular}
              }
            }
            child {
              node [transferred] {%
                \begin{tabular}{c}
                  \ChanNothing\\\ChanNothing
                  \\\ChanPos
                  \\[1.0ex]\LARGE v : Val G12S
                \end{tabular}
              }
            }
          }          
          child {
            node {%
            \begin{tabular}{c}
              \ChanNothing\\\ChanNothing\\\ChanNothing
              \\[1.0ex]\LARGE kR : Cont G2R
            \end{tabular}
            }
          }
        }
        child [sibling distance = 8cm] {
          node [affected] {%
            \begin{tabular}{c}
              \ChanNothing\\\ChanPosUpdated
              \\\ChanNothing
              \\[1.0ex]\LARGE \AgdaInductiveConstructor{Ready} : \AgdaDatatype{Command}\AgdaSpace\AgdaBound{G₁₂'}
            \end{tabular}
          }
          %%%
          child {
            node {%
              \begin{tabular}{c}
                \ChanNothing\\\ChanPosUpdated\\\ChanNothing
                \\[1.0ex]\LARGE vchS : Val G1S
              \end{tabular}
            }
          }
          child {
            node {%
              \begin{tabular}{c}
                \ChanNothing\\\ChanNothing
                \\\ChanNothing
                \\[1.0ex]\LARGE kS : Cont G2S
              \end{tabular}
            }
          }
        }
      }
      child {
          node  {%
            \begin{tabular}{c}
              \ChanPosNeg\\\ChanNothing\\\ChanNeg
              \\[1.0ex]\LARGE \AgdaBound{G₂} --- remaining threads
            \end{tabular}
          }
      }
      ;
    \end{tikzpicture}

%%% Local Variables: 
%%% mode: latex
%%% TeX-master: "ppdp19-mechanized"
%%% End: 

%% file: pic-resource-splitting-values.tex
    \begin{tikzpicture}[->,>=stealth',level/.style={sibling distance = 5cm,
        level distance = 3cm},thick]
      \node {%
        \begin{tabular}{c}
          \ChanNeg\\\ChanPos\\\ChanNothing
          \\Pair (Value 2)
        \end{tabular}
      }
      child {
        node {%
          \begin{tabular}{c}
            \ChanNothing\\\ChanPos\\\ChanNothing
            \\Pair
          \end{tabular}
        }
        child {
          node  {%
            \begin{tabular}{c}
              \ChanNothing\\\ChanPos\\\ChanNothing
              \\Channel
            \end{tabular}
          }
        }
        child {
          node  {%
            \begin{tabular}{c}
              \ChanNothing\\\ChanNothing
              \\\ChanNothing
              \\Int
            \end{tabular}
          }
        }
      }
      child {
          node  {%
            \begin{tabular}{c}
              \ChanNeg\\\ChanNothing\\\ChanNothing
              \\Channel
            \end{tabular}
          }
      }
      ;
    \end{tikzpicture}

%%% Local Variables: 
%%% mode: latex
%%% TeX-master: "popl19-defint"
%%% End: 